\documentclass[11pt,preprint,showpacs,preprintnumbers,amsmath,amssymb,showkeys,floatfix]{revtex4}
\pdfoutput=1

\usepackage{graphicx}
\usepackage{tikz}
\usepackage{dcolumn}
\usepackage{bm}
\usepackage{color} 
\usepackage{url} 

\graphicspath{{./}}

\begin{document}

\preprint{APS/JCP-FMP2}

\title{Tensor factorizations of local second-order M\o ller Plesset theory}
 
\author{Jun Yang}
 \email{jy459@cornell.edu} 
\affiliation{Department of Chemistry and Chemical Biology \\ Cornell University \\ Ithaca, New York 14853, USA}

\author{Yuki Kurashige}
 \email{kura@ims.ac.jp} 
\affiliation{Department of Theoretical and Computational Molecular Science \\ Institute for Molecular Science 
              \\ 38 Nishigo-Naka, Myodaiji, Okazaki 444-8585, Japan}

\author{Frederick R. Manby}%
 \email{fred.manby@bris.ac.uk} 
\affiliation{Center for Computational Chemistry \\ School of Chemistry \\ University of Bristol \\ Bristol BS8 1TS, United Kingdom}

\author{Garnet K. L. Chan}%
 \email{gc238@cornell.edu} 
\affiliation{Department of Chemistry and Chemical Biology \\ Cornell University \\ Ithaca, New York 14853, USA }

\date{\today}

\begin{abstract}

Efficient electronic structure methods can be built around efficient 
tensor representations of the wavefunction. Here we describe
a general view of tensor factorization for the compact representation of electronic wavefunctions.
We use these ideas to construct low-complexity representations of the doubles amplitudes in local second order
M\o{}ller-Plesset perturbation theory. We introduce two approximations - the direct
orbital specific virtual approximation and the full orbital specific virtual approximation. 
In these approximations, each occupied orbital is associated with a small set of correlating virtual orbitals.
Conceptually, the representation lies  between the projected atomic orbital representation  in  Pulay-Saeb\o{}
local correlation theories and  pair natural orbital correlation theories. We have tested
the orbital specific virtual approximations on a variety of systems and properties  including total energies, reaction energies, and potential
energy curves. Compared to the Pulay-Saeb\o{} ansatz, we find that these approximations exhibit favourable accuracy and computational times,
while yielding smooth potential energy curves. 

\end{abstract}

\maketitle

\section{Introduction}
   In electron correlation there are two problems of complexity. The first relates to the information (storage)
required to represent the wavefunction and the second to the complexity of manipulating the wavefunction
to calculate observables. Consider, for example, the doubles amplitudes $t_{ij}^{ab}$ (where  $ij$ denote
occupied orbitals, $ab$ virtual orbitals) common to the second-order M\o{}ller-Plesset 
perturbation (MP2), coupled cluster doubles (CCD), and coupled electron pair approximations (CEPA).
 The storage scales  like $N^4$ while the cost of obtaining the  energy 
scales like $N^5$ for MP2 and $N^6$ for CCD and CEPA, where $N$ is a measure of the size of the system. 

In several limits, we expect these formal complexities to be too high.
For example, if there is a large number of atomic orbitals on a single center, there
is redundancy in the product occupied-virtual pair basis.
Also, in  large molecules 
the cost to obtain the energy should be linear in the size of the molecule.
In both these situations, the mismatch between  formal complexity and our expected  complexity
suggests that the amplitudes and amplitude equations have some special structure.
For example, in Pulay-Saeb{\o} local correlation theories based on the projected atomic orbital (PAO) 
ansatz~\cite{pulay1,pulay2,pulay3} 
the sparsity structure of the amplitudes is built in through distance-based truncations~\cite{BP},
and using this framework it has been possible to achieve linear scaling of storage 
and computational cost with system size~\cite{lccsd_1,lmp2,lccsd_2,lccsdt,discpao2}.
Naturally, these truncations are not orbitally invariant and require
 a representation of the orbitals in which the amplitude matrix is maximally sparse.
An important task in devising algorithms with reduced complexities is to find optimal transformations
of the amplitudes to  representations that are approximable with low complexity. 

The doubles amplitude $t_{ij}^{ab}$ is a tensorial quantity. Consequently, the problem of finding a low
complexity  representation can be viewed as one of tensor representation or tensor factorization. While there has been much recent work
in constructing low complexity representations for the doubles amplitudes and two-electron integrals,
both through using  more optimal orbitals (for example, along the lines
of pair natural orbitals (PNO)~\cite{pnoci1,pnoci2,pnoci3,pnocicepa,pnocepa0,pnocepa,pnoccsd}, 
optimized virtual orbital spaces~\cite{ovos1,ovos2}, 
frozen natural orbitals~\cite{davidson,fnocc1,fnocc2,eomipcc}, and others~\cite{nlmos1, nlmos2}) as well 
as matrix factorizations of the integrals and amplitudes 
(such as Cholesky decompositions (CD) and density fitting (DF) (or resolutions of the identity, RI)
~\cite{df1,df2,df3,df4,df5,df6,df7,df8,df9,dfmp2,dfr12int}), 
these approaches have not yet explored the full generality of the tensorial structure.
Exploring the possibility of a more general class of tensor factorizations is the concern of the current work.

Here we first start with a general  classification of the different kinds of tensor factorizations.
Next we introduce a specific factorization
which we refer to as an orbital specific virtual approximation. This representation
has a simple and intuitive interpretation that bridges
 earlier work on optimal virtuals and work on pair natural orbitals.
We explore the orbital specific virtual approximation in the context of local second order M\o{}ller-Plesset
perturbation theory.  We present benchmark applications on a variety of large molecules, clusters, and reactions. 
We find that the ansatz is very favourable both in its
formal properties such as potential energy curve smoothness, and weak computational dependence on the size of the underlying basis, as well
as regarding its absolute costs in terms of storage and timings when compared to an existing efficient implementation 
of the local Pulay-Saeb{\o} correlation ansatz \cite{molpro,lmp2}.
Finally, we finish with a discussion of the future prospects of such an approach.

\section{Theory and algorithm}
   \subsection{Classification of tensor factorizations}

\begin{figure*}[ht!]
 {\includegraphics[width=0.5\linewidth]{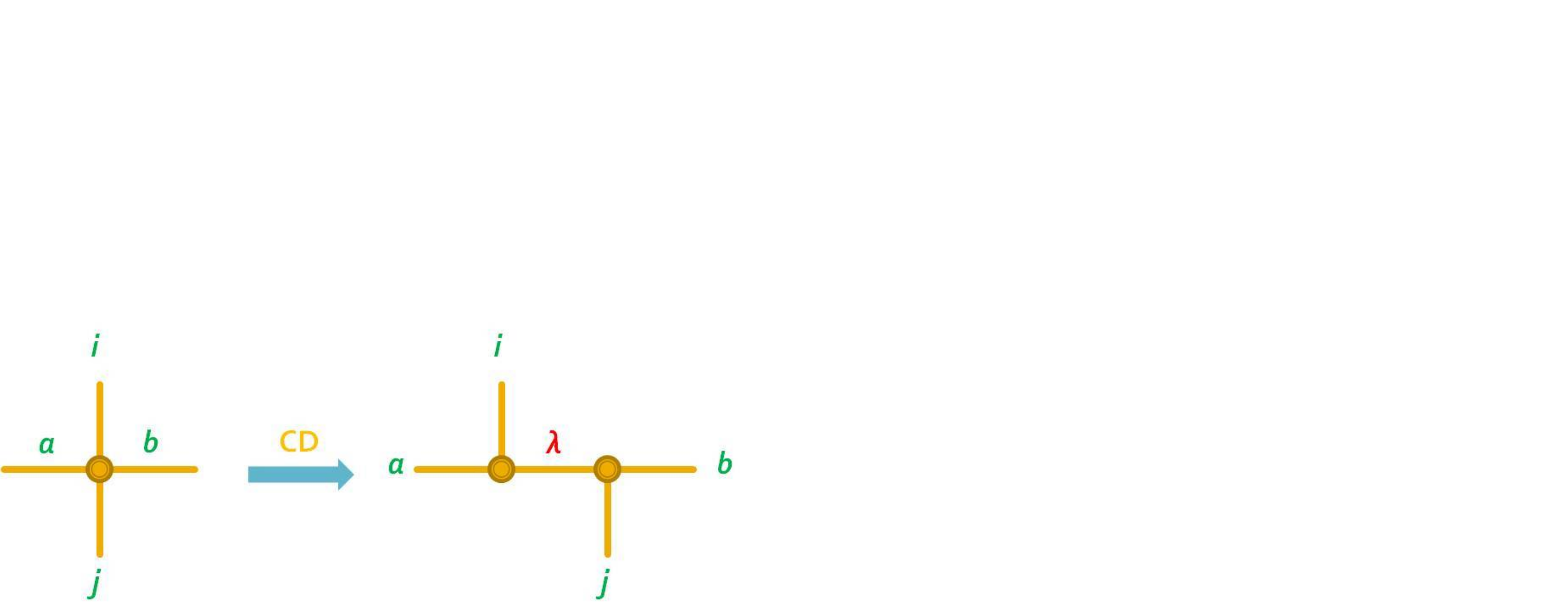}} \\ (a) Cholesky Decomposition \\[1cm]
 {\includegraphics[width=0.5\linewidth]{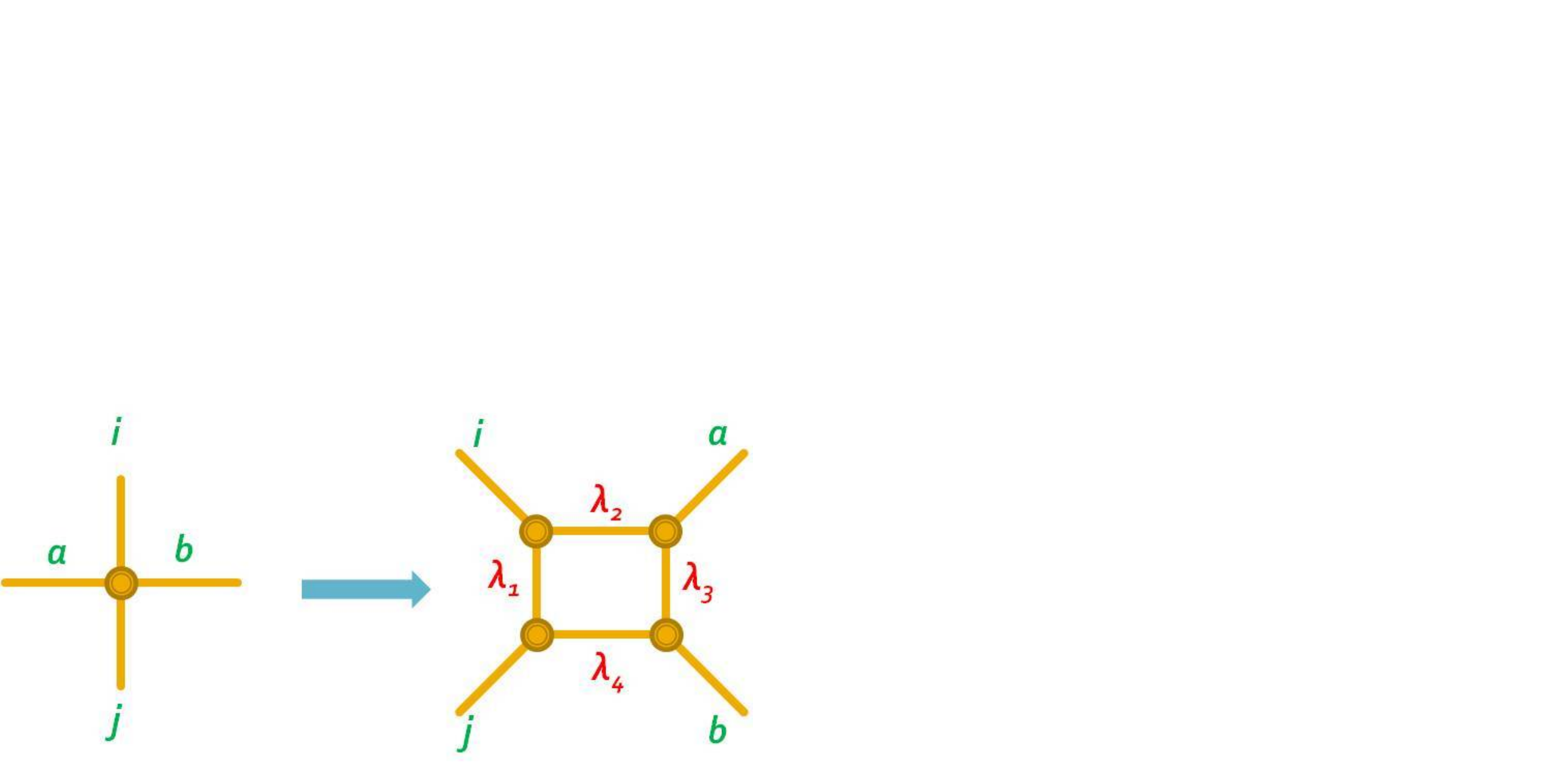}} \\ (b) Matrix Product Decomposition 
 \caption{\label{fig:cd} 
          Pictorial representations of (a) Cholesky Decomposition and (b) Matrix Product Factorization.
        } 
\end{figure*}

\begin{figure*}[ht!]
 {\includegraphics[width=0.5\linewidth]{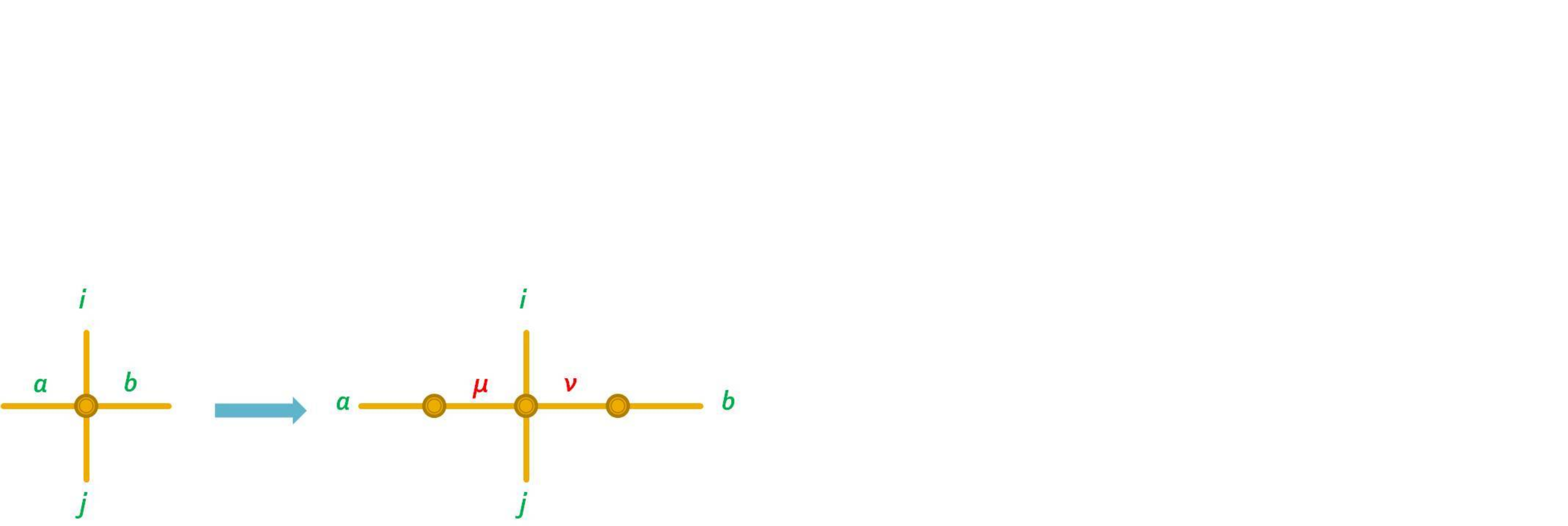}} \\ (a) Pulay-Saeb{\o} PAO      \\[1cm]
 {\includegraphics[width=0.5\linewidth]{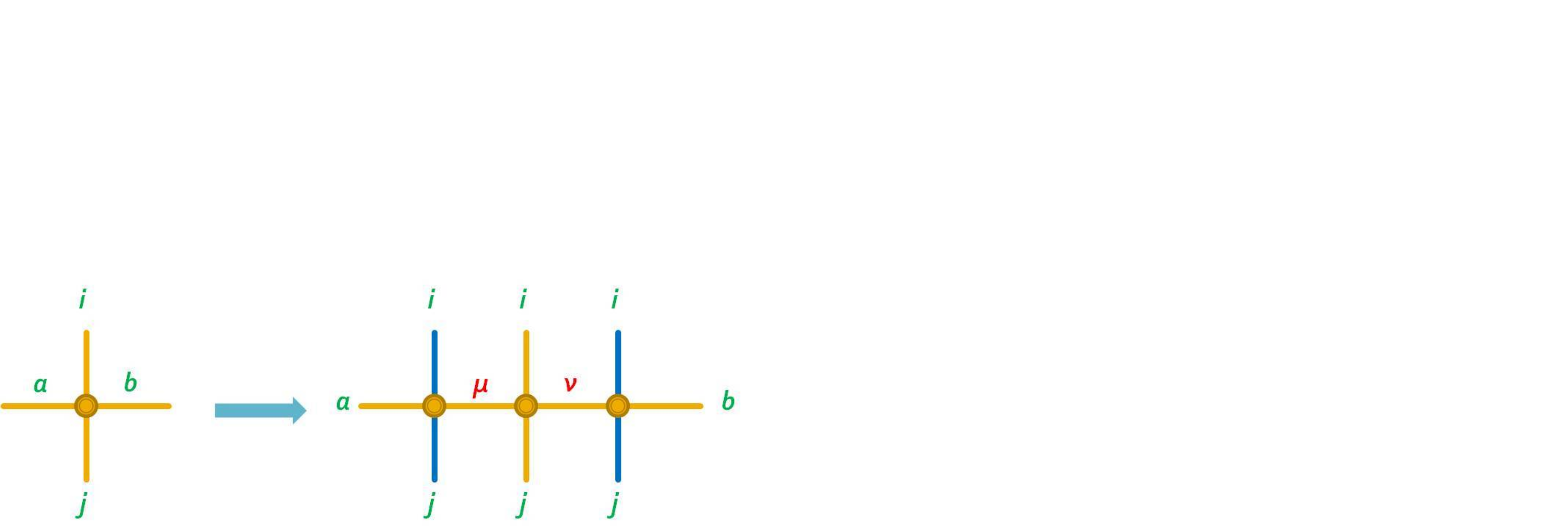}} \\ (b) Pair Natural Orbital    \\[1cm]
 {\includegraphics[width=0.5\linewidth]{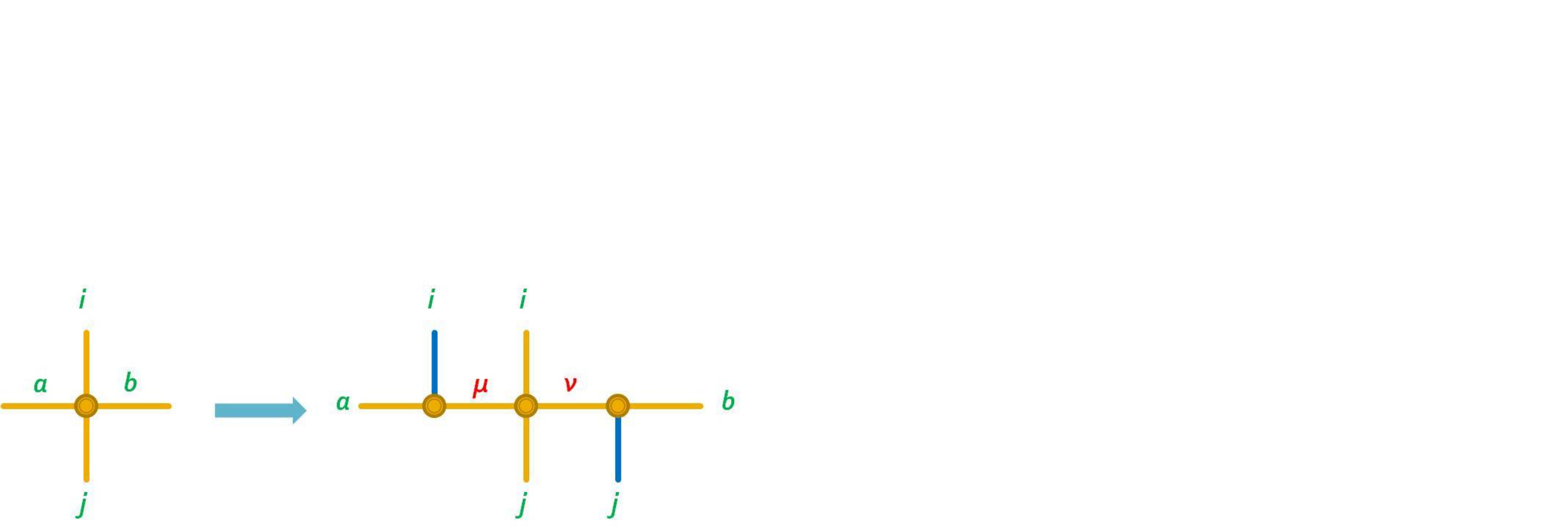}} \\ (c) Orbital Specific Virtual
\caption{\label{fig:facforms} 
         Pictorial representations of (a) $t_{ij}^{ab} = \sum_{\mu \nu} t_{ij}^{\mu\nu} t_{\mu}^a  t_{\nu}^b$ (Pulay-Saeb{\o} PAO approximation),
         (b) $t_{ij}^{ab}=\sum_{\mu\nu}t_{ija}^{\mu_{ij}}t^{\mu_i\nu_j}_{ij}t_{ijb}^{\nu_{ij}}$ (Pair Natural Orbital approximation) and  
         (c) $t_{ij}^{ab}=\sum_{\mu\nu}t_{ia}^{\mu_i}t^{\mu_i\nu_j}_{ij}t_{jb}^{\nu_j}$ (Orbital Specific Virtual approximation).
        }
\end{figure*}

We are concerned primarily with the doubles amplitude tensor $t_{ij}^{ab}$. We illustrate it pictorially
as a connected four-point object (see objects on the left in Figs. \ref{fig:cd} and~\ref{fig:facforms}). A closely
related quantity, particularly in  second-order M\o{}ller-Plesset theory, is the two-electron integral $v_{ij}^{ab}$. In canonical
closed-shell MP2 theory, the two are related by
\begin{align}
t_{ij}^{ab} = (2 v_{ij}^{ab} - v_{ij}^{ba})(\epsilon_i + \epsilon_j - \epsilon_a - \epsilon_b)^{-1}
\end{align}

To construct a low-complexity representation of the two-electron integrals or amplitudes, we
   approximate the high-dimensional amplitude or integral tensors by lower-dimensional components. These
component tensors may share the same ``physical'' indices $i,j,a,b$
 as the target tensor, but may  also carry additional ``auxiliary'' indices $\lambda, \mu, \nu, \rho, \ldots$.
Since the auxiliary indices do not appear in the target tensor,
they must be traced over in some way, and both the distribution of
the physical and auxiliary indices amongst the component tensors, as well as the pattern of contractions,
 defines the particular tensor representation. Since these contractions are usually non-linear, it is often
useful to visualize the contractions pictorially rather than algebraically. 

To illustrate this, consider first the
 density-fitting and Cholesky decomposition approximations.
In all these approximations, the two-electron integrals are viewed as a matrix factorization as 
\begin{align}
v_{ij}^{ab} = \sum_{\lambda} L_{i\lambda}^a L_{j\lambda}^b
\end{align}
where $\lambda$ is the auxiliary index. Pictorially, we  view
 the above as separating the $ia$, $jb$ electron-hole degrees of freedom, which must then be reconnected
 via an  auxiliary index (see Fig.~\ref{fig:cd}a). Naturally, the efficiency of the factorization
relies on the rank of the decomposition (the number of terms in the sum) being low.
As another example, consider the types of correlation ansatz (such as the Pulay-Saeb{\o} local correlation ansatz)
which use a non-canonical virtual orbital basis, for example the projected atomic orbital (PAO) virtuals.
Non-canonical virtuals $\phi_\mu$ are  related to  canonical virtuals $\phi_a$ by a transformation
\begin{align}
\phi_\mu = \sum_{\mu} t_{\mu}^a \phi_a
\end{align}
and consequently, the canonical and non-canonical doubles amplitudes are related by
\begin{align}
t_{ij}^{ab} = \sum_{\mu \nu} t_{ij}^{\mu\nu} t_{\mu}^a  t_{\nu}^b \label{eq:pao_ansatz}
\end{align}
which defines the amplitude approximation. Pictorially, this approximation is illustrated in Fig.~\ref{fig:facforms}a. 
For appropriate $t_\mu^a$ (such as defined by the PAO virtuals), representation (\ref{eq:pao_ansatz}) allows one to favourably exploit locality.
For example, in the Pulay-Saeb\o{}-Werner-Sch\"utz approaches~\cite{pulay1,pulay2,pulay3,lmp2,lccsd_2,lccsdt,discpao2}, 
a sparsity structure on $t_{ij}^{\mu\nu}$ is imposed by requiring $t_{ij}^{\mu\nu}=0$ when $i$ is far apart from $j$,
and for other $ij$, the sum over $\mu$ and $\nu$ is restricted to defined  domains $[ij]$ that are in the spatial vicinity of $ij$. 

Thus, the essence of low-complexity tensor approximation is captured by the 
types of indices on the components and their connectivity, as illustrated
in their pictorial representation. We can 
consider  generalizations of the above approximations in a variety of ways. For example, we can 
consider approximations with additional
auxiliary indices. One example is (cf. Fig.~\ref{fig:cd}b)
\begin{align}
t_{ij}^{ab} &= \sum_{\lambda_1 \ldots \lambda_4} t_{i}^{\lambda_1 \lambda_2} t^{a}_{\lambda_2 \lambda_3} t^b_{\lambda_3 \lambda_4} t_j^{\lambda_4 \lambda_1} \\
&= \mathrm{tr} [\mathbf{t}_i \mathbf{t}^a \mathbf{t}^b\mathbf{t}_j]  \label{eq:mps_factor}
\end{align}
where the amplitude $t_{ij}^{ab}$ is reconstructed as a trace of a matrix product.
This  approximation recalls the matrix product factorization  in the density matrix
renormalization group (DMRG)~\cite{dmrg1,dmrg2}, although the physical content here is quite distinct, 
since the  DMRG is carried out in the occupation number space rather than in the excitation space. 
Another way to construct new approximations is to introduce components
with repeated physical indices. A particularly simple example is 
\begin{align}
t_{ij}^{ab} = t_{ij} t_{ia} t_{jb} 
\end{align}
where there are no auxiliary indices at all.  
This recalls the correlator product state approximation (also known as an entangled plaquette state~\cite{eps1,eps2}), although once again the tensor is expressed in an excitation rather than occupation number picture.
Naturally, we can consider many other combinations of  auxiliary indices and physical indices,
and the appropriateness of the particular choice depends on the problem at hand.

\subsection{Orbital specific virtual approximation}

We now consider a simple tensor factorization of the doubles amplitudes $t_{ij}^{ab}$ 
that we will study in this work. 
We first define the \textit{direct orbital specific virtual}~(dOSV) approximation to the amplitudes as (cf. Fig.~\ref{fig:facforms}c)
\begin{align}
 t_{ij}^{ab}=\sum_{\mu\nu}t_{ia}^{\mu_i}t^{\mu_i\nu_j}_{ij}t_{jb}^{\nu_j},\label{eqn:fac}
\end{align}
This has a simple physical interpretation: the component $t_{ia}^{\mu_i}$ (and similarly $t_{jb}^{\nu_j}$) defines
a set of virtual orbitals for \textit{each} occupied orbital, and
 $t_{ij}^{\mu_i\nu_j}$ represents amplitudes in this orbital specific basis. 
Note that the subscript $i$ in $\mu_i$ is somewhat redundant, but we retain it to emphasize that $\mu_i$ labels an orbital specific virtual 
associated with occupied orbital $i$.
By choosing a good set of components $t_{ia}^{\mu_i}$, either by direct optimization or otherwise (see later) 
we may define suitable adaptations of the
virtual basis for each occupied orbital. This is quite natural
in a local correlation theory, as the optimal orbital specific virtuals for a 
localized occupied orbital must be located in close spatial proximity; however, 
even when the occupied orbitals are delocalized, we can still expect
this factorization to be beneficial, as a given occupied orbital does
not correlate equally with all parts of the virtual space.

In the direct orbital specific virtual approximation, occupied orbital $i$ excites only to its virtual set $\mu_i$ ($i\to\mu_i$),
and occupied orbital $j$ only to its orbital set $\nu_j$ ($j\to\nu_j$), the ``exchange'' excitations $i \to \nu_j, j \to \mu_i$ being
excluded. 
It was shown, however,  in the context of Pulay-Saeb{\o} local theory,
 that the inclusion of exchange excitations can lead to greatly improved results~\cite{cross0,cross1, cross2, discpao1}.
While formally the exchange excitations can be included by increasing the size of sets $\mu_i$ and $\nu_j$, we can also include
 them explicitly in the structure of the ansatz, which leads to the full orbital specific  virtual (OSV) approximation 
\begin{align}
t_{ij}^{ab}=\sum_{\mu\nu} \left(t_{ia}^{\mu_i} t_{ij}^{\mu_i \nu_i} t_{jb}^{\nu_i}
+  t_{ia}^{\mu_i} t_{ij}^{\mu_i \nu_j} t_{jb}^{\nu_j}
+  t_{ia}^{\mu_j} t_{ij}^{\mu_j \nu_i} t_{jb}^{\nu_i}
+  t_{ia}^{\mu_j} t_{ij}^{\mu_j \nu_j} t_{jb}^{\nu_j}\right)
\end{align}
This can be written in matrix form
\begin{align}
t_{ij}^{ab}=\sum_{\mu\nu} \left(\begin{array}{cc} t_{ia}^{\mu_i} & t_{ia}^{\mu_j} \end{array}\right)
\left(\begin{array}{cc} t_{ij}^{\mu_i \nu_i} & t_{ij}^{\mu_i \nu_j}\\ 
                        t_{ij}^{\mu_j \nu_i} & t_{ij}^{\mu_j \nu_j}\end{array}\right) 
 \left(\begin{array}{c} t_{jb}^{\nu_i} \\ t_{jb}^{\nu_j} \end{array}\right)
\end{align}

To gain further understanding, we  briefly discuss
the connection to  other approximations where
 non-canonical virtual spaces are used. In theories
which use a global set of non-canonical virtuals, such 
as the projected atomic orbital virtual space in the Pulay-Saeb{\o} ansatz,
the amplitude tensor is expressible as in Eq. (\ref{eq:pao_ansatz}), 
where $t^a_\mu$ and $t^b_\nu$  parametrize the transformation to a new virtual basis. In pair
natural orbital theories  pioneered some time ago~\cite{pnoci1,pnoci2,pnoci3,pnocicepa,pnocepa0} 
and which have been recently revisited by Neese and coworkers~\cite{pnocepa, pnoccsd}, 
 a correlating virtual space is
defined for each occupied pair $ij$, and  the amplitudes are factorized as (cf. Fig.~\ref{fig:facforms}b)
\begin{align}
 t_{ij}^{ab}=\sum_{\mu\nu}t_{ija}^{\mu_{ij}}t^{\mu_i\nu_j}_{ij}t_{ijb}^{\nu_{ij}}.\label{eqn:facP}
\end{align}
Compared to the Pulay-Saeb{\o} ansatz where
a global set of virtuals is used, the orbital specific virtual approximation
is able to adapt the virtual space, which leads to a more compact representation of the amplitudes. On the
other hand, the orbital specific virtual approximation adapts each virtual space to a single occupied
orbital rather than a pair. This avoids some of the complexities inherent
to the pair natural orbital ansatz where the definitions of the virtual spaces 
involve four-index components $t_{ija}^{\mu_{ij}}$ of similar formal complexity to the amplitudes themselves, and which lead
to complicated overlap and Fock matrices in the pair natural orbital virtual blocks. 
Consequently, we see that formally the orbital specific virtual approximation interpolates between the
Pulay-Saeb{\o} form and the pair natural orbital approximation. We now turn towards its practical implementation in second-order perturbation theory (MP2).


\section{Implementation}
   \subsection{MP2 wavefunction and singular value orbital specific virtuals}

The central task of MP2 theory is to determine the first-order wavefunction $|\Psi^{(1)}\rangle$,
\begin{align}
|\Psi^{(1)}\rangle = \frac{1}{2}\sum_{ijab} t_{ij}^{ab} |\Phi_{ij}^{ab}\rangle \label{eqn:1stmp2}
\end{align}
where in 
the above  $i,j,\cdots$ and $a,b,\cdots$ refer respectively to the occupied and virtual \textit{spatial} orbitals.
Explicitly in spin-orbital notation
\begin{align}
|\Phi_{ij}^{ab}\rangle = \sum_{\sigma,\sigma^\prime \in \{ \alpha,\beta\}} |\Phi_{i\sigma j\sigma^\prime}^{a\sigma b\sigma^\prime}\rangle 
\end{align}
For closed-shell molecules the spin-free orbital notation avoids the explicit use of spin coordinates and is very convenient.

Inserting the orbital specific virtual approximation (cf. Eq.~(\ref{eqn:fac})) into Eq.~(\ref{eqn:1stmp2}),
we  parametrize the first-order wavefunction in terms of factorized amplitudes $t_{ij}^{\mu_i\nu_j}$
and corresponding determinants $| \Phi^{\mu_i\nu_j}_{ij} \rangle$. For the direct orbital specific virtual approximation of MP2 (dOSVMP2),
\begin{align}
   | \Psi^{(1)} \rangle = \frac{1}{2}\sum_{ij\mu\nu} t_{ij}^{\mu_i\nu_j}| \Phi^{\mu_i\nu_j}_{ij} \rangle \label{eqn:dosvmp2}
\end{align}
where 
\begin{align}
  |\Phi^{\mu_i\nu_j}_{ij} \rangle = \sum_{ab} t_{ia}^{\mu_i} |\Phi_{ij}^{ab}\rangle t_{jb}^{\nu_j}
\end{align}
while for the full orbital specific virtual approximation of MP2 (OSVMP2) we have
\begin{align}
   | \Psi^{(1)} \rangle = \frac{1}{2}\sum_{ij\mu\nu} \left(t_{ij}^{\mu_i\nu_i}| \Phi^{\mu_i\nu_i}_{ij}\rangle
+t_{ij}^{\mu_i\nu_j}| \Phi^{\mu_i\nu_j}_{ij}\rangle
+t_{ij}^{\mu_j\nu_i}| \Phi^{\mu_j\nu_i}_{ij}\rangle
+t_{ij}^{\mu_j\nu_j}| \Phi^{\mu_j\nu_j}_{ij}\rangle\right)  \label{eqn:osvmp2}
\end{align}
In Eq. (\ref{eqn:dosvmp2}) we have introduced an orbital specific virtual $| \mu_i \rangle$ (OSV) defined through an orbital transformation 
from the  virtual $| a \rangle$ using $t_{ia}^{\mu_i}$,
\begin{align}
 | \mu_i \rangle = \sum_a t_{ia}^{\mu_i} | a \rangle , \label{eqn:osv}
\end{align}

Naturally, we would like the OSVs to be well adapted to each occupied orbital in Eq.~(\ref{eqn:osv}).
One quick and economical scheme to determine $t_{ia}^{\mu_i}$ is to perform a singular value decomposition (SVD) 
of the MP2 diagonal amplitudes $t_{ii}^{ab}$ for \textit{each} occupied orbital $i$, 
\begin{eqnarray}
  t_{ii}^{ab} & = & \sum_\mu t_{ia}^{\mu_i} s_{\mu_i} t_{ib}^{\mu_i}. \label{eqn:tau}
\end{eqnarray}
where $s_{\mu_i}$ is the singular value.
In the canonical basis, $t_{ii}^{ab}$ is directly calculated from,
\begin{align}
  t_{ii}^{ab} = \frac{v_{ii}^{ab}}{2\epsilon_i - \epsilon_a - \epsilon_b}.
\end{align}
When the localized occupied orbitals are used, $\epsilon_i$ is the diagonal element of the Fock matrix in the local orbital basis.
$\epsilon_a$ and $\epsilon_b$ are the diagonal elements of the virtual block of Fock matrix.

The SVD provides a natural setting to truncate the OSV space, as the singular vectors with small singular values $s_\mu$
should contribute little to the final amplitudes. 
Consequently it is reasonable to include only those vectors with the largest eigenvalues,
keeping either a fixed number of OSVs per occupied orbital, a fixed percentage of OSVs, or by using a numerical threshold on $s_\mu$. We have used
the first two truncation schemes in this work. After truncation, the complete virtual space is parametrized by an incomplete set of orbital specific
virtuals. 
This of course introduces errors relative to canonical MP2 theory.
However, as numerically shown in the next section, 
the resulting correlation energies exhibit only minor deviations (e.g. $<0.01\%$) from canonical values, while achieving
very substantial gains in computational efficiency.

The OSVs defined above are not always orthogonal. The overlap matrix between the OSVs  $|\mu_i \rangle$ and $|\nu_j \rangle$ is,
\begin{align} 
  S_{\mu_i\nu_j} = \langle \mu_i | \nu_j \rangle  = \sum_a t_{ia}^{\mu_i} t_{ja}^{\nu_j}
\end{align}
Through the SVD, the OSVs $|\mu_i \rangle$ belonging to the same occupied orbital are orthonormal 
but the OSVs from  different occupied orbitals are not.

We also note that the  SVD does not necessarily  yield the most optimal orbital specific virtual orbitals. 
The direct optimization of $t_{ia}^{\mu_i}$ relaxes the OSVs and may help achieve a more compact description of the correlation effects. 
This  is under investigation and will be presented elsewhere.

\subsection{Residual equations}
We derive the exact MP2 residual equations in the orbital specific virtual basis starting from the Hylleraas functional,
\begin{align}
  h = \langle \Psi^{(1)} | \textrm{F} - E^{(0)} | \Psi^{(1)} \rangle 
            + 2 \langle \Psi^{(1)} | \textrm{V} | \Psi^{(0)} \rangle, \label{eqn:hyll}
\end{align}
where $\textrm{F}$ is the Fock operator and $\textrm{V}$ is the two-electron fluctuation potential, respectively.
$E^{(0)}$ is the sum of occupied Hartree-Fock eigenvalues.
By parametrizing $\Psi^{(1)}$ in the direct orbital specific virtual approximation (dOSVMP2, cf. Eq.~(\ref{eqn:dosvmp2})) 
and making the first derivative of $h$ with respect to $t_{ij}^{\mu_i\nu_j}$ vanish,
we arrive at the following formal residual,
\begin{align}
 R_{ij}^{\mu_i\nu_j} = \left( \frac{\partial h~~~}{\partial t_{ij}^{\mu_i\nu_j}}  \right)
                     = \langle \Phi_{ij}^{\mu_i\nu_j} | \textrm{F} - E^{(0)} | \Psi^{(1)} \rangle
                       + \langle \Phi_{ij}^{\mu_i\nu_j} |\textrm{V} | \Psi^{(0)} \rangle
                     = 0. \label{eqn:res1}
\end{align}
The expansion of $R_{ij}^{\mu_i\nu_j}$ has the following explicit form for a particular pair $(i,j)$,
\begin{widetext}
\begin{align}
  \mathrm{R}_{(i,j)} =& 2\mathrm{K}_{(i,j)} - \mathrm{C}_{(i,j)} +
                   2\left[\mathrm{T}_{(i,j)}\mathrm{F}_{(j,j)} + \mathrm{F}_{(i,i)}\mathrm{T}_{(i,j)} 
                            - \sum_k F_{kj}\mathrm{T}_{(i,k)}\mathrm{S}_{(k,j)} - \sum_k F_{ik} \mathrm{S}_{(i,k)}\mathrm{T}_{(k,j)}\right] \nonumber \\
                & - \mathrm{S}_{(i,j)}\left[\mathrm{T}_{(j,i)}\mathrm{F}_{(i,j)} - \sum_k F_{ik} \mathrm{T}_{(j,k)}\mathrm{S}_{(k,j)}\right]
                 - \left[\mathrm{F}_{(i,j)}\mathrm{T}_{(j,i)} - \sum_k F_{kj}\mathrm{S}_{(i,k)}\mathrm{T}_{(k,i)}\right]\mathrm{S}_{(i,j)}.
                          \label{eqn:res2}
\end{align}
\end{widetext}
Here $\mathrm{T}_{(i,j)}$ is the matrix form of amplitudes $\{t_{ij}^{\mu_i\nu_j}\}$, fixing $i,j$.
$F_{ik}$ and $F_{kj}$ are elements of the occupied block of the Fock matrix.
$\mathrm{K}_{(i,j)}$ and $\mathrm{C}_{(i,j)}$ denote the matrices storing two-electron Coulomb and exchange integrals 
 $(\phi_i\phi_{\mu_i}|\phi_j\phi_{\nu_j})$ and $(\phi_i\phi_{\nu_j}|\phi_j\phi_{\mu_i})$, respectively.
$\mathrm{S}_{(i,j)}$ and $\mathrm{F}_{(i,j)}$ are, respectively, the overlap and Fock matrices  with elements
 $\langle \phi_{\mu_i}|\phi_{\nu_j}\rangle$ and $\langle \phi_{\mu_i}|\hat{F}|\phi_{\nu_j}\rangle$
for a pair $(i,j)$.
None of the above quantities gives rise to real bottlenecks for memory or disk storage
if the orbital specific virtual space is  truncated.

In the full orbital specific virtual approximation (OSVMP2, cf. Eq.~(\ref{eqn:osvmp2})), 
we minimize the Hylleraas functional with respect to the amplitudes $t_{ij}^{\mu_i\nu_i}, 
t_{ij}^{\mu_i\nu_j}, t_{ij}^{\mu_j\nu_i}, t_{ij}^{\mu_j\nu_j}$ giving rise to analogues of (\ref{eqn:res1}). The explicit residuals are then 
\begin{align}
 \mathrm{R}_{(i,j)} = \mathrm{K}_{(i,j)}\!+\!
  \sum_{k}\!\left\{\mathrm{S}_{(ij,ik)}\mathrm{T}_{(i,k)}\left[ \delta_{kj} \mathrm{F}_{(ik,ij)} - F_{kj} \mathrm{S}_{(ik,ij)} \right] 
  \!+\!\left[\delta_{ik}\mathrm{F}_{(ij,kj)}-F_{ik}\mathrm{S}_{(ij,kj)} \right]\mathrm{T}_{kj} \mathrm{S}_{(kj,ij)}\right\}. \label{eqn:resC}
\end{align}
where $\mathrm{R}_{(i,j)}$ is now a matrix of dimension $\textrm{dim}(\mu_i)+\textrm{dim}(\nu_j)$, with
elements of types $R_{ij}^{\mu_i\nu_i}, R_{ij}^{\mu_i\nu_j}, R_{ij}^{\mu_j\nu_i}, R_{ij}^{\mu_j\nu_j}$, 
 and similarly for $\mathrm{K}_{(i,j)}$. 
 $\mathrm{S}_{(ij,ik)}$ is the overlap matrix between
the bras $\{ \langle \mu_i|, \langle \mu_j| \}$ and  the kets $\{ |\nu_i\rangle, |\nu_k\rangle \}$,
and  $\mathrm{F}_{(ij,ik)}$ is the analogously defined Fock matrix.

\subsection{Projective residual equations for  dOSVMP2 (dOSVMP2-P)}

Compared to the standard local MP2 residual equations, those for the orbital specific virtuals appear more complicated,
especially for the direct (dOSVMP2) ansatz.
Formally, we can obtain the MP2 amplitudes not only 
through variational minimization of the Hylleraas functional, but
also by projection with an appropriate set of bra states. In the case
of dOSVMP2, this leads to a simpler set
of residual equations which define a different set of amplitudes than those arising from Eq. (\ref{eqn:res2}).
We write
\begin{align}
 \widetilde{R}_{ij}^{\mu_i\nu_j} = \langle \widetilde{\Phi}_{ij}^{\mu_i\nu_j} | \textrm{F} - E^{(0)} | \Psi^{(1)} \rangle
                       + \langle \widetilde{\Phi}_{ij}^{\mu_i\nu_j} |\textrm{V} | \Psi^{(0)} \rangle
                     = 0. \label{eqn:resP1}
\end{align}
The bra states are chosen to be bi-orthonormal to the ket states in the spatial orbital basis following Ref.~\cite{biortho,pulay2},
\begin{align}
\langle \widetilde{\Phi}^{\mu_i\nu_j}_{ij}|\Phi^{\omega_k\gamma_l}_{kl} \rangle 
                = \frac{1}{\sqrt{1+\delta_{ij}\delta_{\mu_i\nu_j}}}\left(\delta_{ik}\delta_{jl}\delta_{\mu\omega}\delta_{\nu\gamma} 
                  + \delta_{jk}\delta_{il}\delta_{\nu\omega}\delta_{\mu\gamma}\right)
\end{align}
with a normalization prefactor.
The bi-orthonormal OSV bra state $\langle \widetilde{\Phi}^{\mu_i\nu_j}_{ij}|$ is defined 
from the canonical bi-orthonormal bra state $\langle \widetilde{\Phi}^{ab}_{ij}|$,
\begin{align}
 \langle \widetilde{\Phi}^{\mu_i\nu_j}_{ij}| = \sum_{ab}\langle \widetilde{\Phi}^{ab}_{ij}|t_{ia}^{\mu_i}t_{jb}^{\nu_j} \label{eqn:biorth}
\end{align}
with~\cite{biortho}
\begin{align}
  \langle \widetilde{\Phi}^{ab}_{ij}| &= \frac{1}{3}\left(2\langle \Phi^{ab}_{ij}| + \langle \Phi^{ba}_{ij}|\right) \nonumber \\
  \langle \widetilde{\Phi}^{ab}_{ij}|\Phi^{cd}_{kl}\rangle &= 
     \frac{1}{\sqrt{1+\delta_{ij}\delta_{ab}}}
        \left(\delta_{ik}\delta_{jl}\delta_{ac}\delta_{bd}+\delta_{jk}\delta_{il}\delta_{bc}\delta_{ad}\right)
\end{align}
Note that if we truncate the OSV space, then the space spanned by the bras $\{ \langle \tilde{\Phi}_{ij}^{\mu_i\nu_j}|\}$ is
not the same as the space spanned by  $\{ \langle {\Phi}_{ij}^{\mu_i\nu_j}|\}$ , due to the presence of the
exchange-like excitation $\langle \Phi_{ij}^{ba}|$ in the definition of $\langle \tilde{\Phi}_{ij}^{ba}|$. This is what gives
rise to the difference between the projective and (standard) Hylleraas based residual equations for the dOSVMP2 approximation.

Thus expanding Eq.~(\ref{eqn:resP1}) yields the projective residual equation $\mathrm{\widetilde{R}}_{(i,j)}$,
\begin{widetext}
\begin{align}
  \mathrm{\widetilde{R}}_{(i,j)} = \mathrm{K}_{(i,j)} 
                            + \mathrm{\widetilde{T}}_{(i,j)}\mathrm{F}_{(j,j)} + \mathrm{F}_{(i,i)}\mathrm{\widetilde{T}}_{(i,j)} 
                            - \sum_k F_{kj}\mathrm{\widetilde{T}}_{(i,k)}\mathrm{S}_{(k,j)} 
                            - \sum_k F_{ik} \mathrm{S}_{(i,k)}\mathrm{\widetilde{T}}_{(k,j)} \label{eqn:resP2}
\end{align}
\end{widetext}
which has clearly a simpler form than the standard residual in Eq.~(\ref{eqn:res2}).
We denote the direct orbital specific virtual approximation defined by the amplitudes from (\ref{eqn:resP2}), the projective
dOSVMP2-P approximation.
In a complete virtual space, the solution of either dOSVMP2 or dOSVMP2-P 
yields the exact canonical MP2 energy and the first-order wavefunction.
In an incomplete virtual space (e.g. if not all the orbital specific virtuals are used),
however, dOSVMP2-P generally does not result in a correlation energy 
that is variationally bounded above the canonical MP2 value with respect to the size of incomplete virtual space. 
The numerical comparison between dOSVMP2 and dOSVMP2-P will be given in the next section.

\subsection{Preconditioning}

As the Fock matrix is diagonal in the canonical basis, the canonical MP2 amplitudes are directly calculated as,
\begin{align}
 t_{ij}^{ab} = \frac{v_{ij}^{ab}}{\epsilon_i+\epsilon_j-\epsilon_a-\epsilon_b},
\end{align}
where $\epsilon$'s are the diagonal elements of the canonical Fock matrix.
However, in the orbital specific virtual space, the Fock matrix contains significant off-diagonal elements and the residual equations must be
solved iteratively. Here the use of a preconditioner is essential.
To this end we define pseudo-virtual energies.
The pseudo-virtual energies can be obtained by
diagonalizing the Fock matrix in the space of orbital specific virtuals for each diagonal occupied pair $(i,i)$, similarly
to as done in Ref.~\cite{lccsd_1}
\begin{align}
 \mathrm{F}_{(i,i)} \mathrm{X}_{(i,i)} = \mathrm{S}_{(i,i)} \mathrm{X}_{(i,i)} \mathrm{\overline{E}}_{(i,i)},
\end{align}
where $\mathrm{X}_{(i,i)}$ is the transformation matrix that diagonalizes both $\mathrm{F}_{(i,i)}$ and $\mathrm{S}_{(i,i)}$.
$\mathrm{\overline{E}}_{(i,i)}$ is diagonal and contains the pseudo-virtual energies.
When preconditioning the residual $\mathrm{R}_{(i,j)}$ corresponding to orbital pair $(i,j)$, we then consider
the residual as a matrix with the first index corresponding to the virtuals associated with $i$,
and the second with the virtuals associated with $j$. 
Consequently, the transformation is performed on the virtual indices of the original residual $\mathrm{R}_{(i,j)}$
with $\mathrm{X}_{(i,i)}^\dag$ and $\mathrm{X}_{(j,j)}$, respectively.
\begin{align}
 \overline{\mathrm{R}}_{(i,j)} =  \mathrm{X}_{(i,i)}^\dag \mathrm{R}_{(i,j)} \mathrm{X}_{(j,j)}.
\end{align}
The amplitude update matrix $\mathrm{\Delta \overline{t}}_{(i,j)}$ in the transformed basis is then,
\begin{align}
 \Delta \overline{t}_{ij}^{\mu_i\nu_j} = \frac{\overline{R}_{ij}^{\mu_i\nu_j}}
                  {\epsilon_i+\epsilon_j-\overline{\epsilon}_{\mu_i}-\overline{\epsilon}_{\nu_j}},
\end{align}
where $\overline{R}_{ij}^{\mu_i\nu_j}$ is the element of the residual matrix $\overline{\mathrm{R}}_{(i,j)}$.
The pseudo-virtual energies of $\overline{\epsilon}_{\mu_i}$ and $\overline{\epsilon}_{\nu_j}$ are the elements of 
 $\mathrm{\overline{E}}_{(i,i)}$ and $\mathrm{\overline{E}}_{(j,j)}$, respectively.
Finally the update in the original orbital specific basis is given by the back-transformation
\begin{align}
 \mathrm{\Delta t}_{(i,j)} = \mathrm{X}_{(i,i)} \mathrm{\Delta \overline{t}}_{(i,j)} \mathrm{X}^{\dagger}_{(j,j)}.
\end{align}
The same algorithm can be used with both dOSVMP2 and OSVMP2, the latter requiring the obvious generalizations
of indices $(i,j)$ to $(i,i)$, $(i,j)$, $(j,i)$, $(j,j)$.

Note that eigenvectors  $\mathrm{X}_{(i,j)}$  may be linearly dependent. (The same issue arises e.g. in the 
Pulay-Saeb{\o} local correlation theory \cite{lccsd_1}). The redundant
vectors are eliminated by a canonical orthogonalization of $\mathrm{S}_{(i,j)}$, discarding eigenvalues below
a given threshold. Here we use a threshold of $10^{-6}$.

With the above preconditioning the amplitudes and MP2 energies  converge very quickly. For example, typically, the change in correlation energy 
falls below $10^{-6}$ a.u. within $8\sim10$ iterations for dOSVMP2-P and OSVMP2
and within $16\sim20$ iterations for dOSVMP2.

\subsection{Computational cost and screening}

The most expensive term in our current implementation of the residual equations~(\ref{eqn:res2}) and~(\ref{eqn:resP2}) 
is the contraction $\sum_k F_{kj}\mathrm{T}_{(i,k)}\mathrm{S}_{(k,j)}$ which has a formal scaling of $O^3V^3$, i.e. $N^6$
with system size.
This is to be contrasted with the canonical MP2 scaling of $N^5$. 
The local Pulay-Saeb{\o} ansatz has a similar contraction, but because there is a single
set of underlying virtual orbitals, the contraction can be implemented in  $O^2V^3 +O^3V^2$ operations, which is still $N^5$ scaling.
In large molecules, $V$ is independent of molecular size, and so the scaling of local OSVMP2 is $O^3$ where
the $O^3$ term has a larger prefactor than in the Pulay-Saeb{\o} theory.

We can  lower the computational cost by introducing some screening approximations to the $N^6$ contraction.
As we have discussed previously, 
the most important OSVs that can be correlated with each occupied orbital are located closely around the occupied orbital itself.
By exploiting  orbital locality, the overlap matrix element $S_{\gamma_k\nu_j}$ decays exponentially 
with the separation of $k$ and $j$. 
Based on this we can simply ignore entire classes of unimportant $N^6$ contractions in the residuals without losing  much  accuracy,
and consequently reduce the formal scaling of solving the residual equations in a large system to order $O^2$. 
The following ratio $t^{\textrm{S}}_{kj}$ for a given pair $(k,j)$ is computed in order to define an appropriate screening threshold, 
\begin{align}
t^{\textrm{S}}_{kj} = \frac{\sum_{\gamma\nu}S^2_{\gamma_k\nu_j}}{\sum_{\gamma\nu}S^2_{\gamma_k\nu_k}}.\label{eqn:screening}
\end{align}
With a screening threshold of $T_{\textrm{S}}$ we then neglect any overlap matrix $\mathrm{S}_{(k,j)}$ 
belonging to a pair $(k,j)$ if $t^{\textrm{S}}_{kj}<T_{\textrm{S}}$.
From the definition $t^{\textrm{S}}_{kj}$ ranges between 0 and 1:
thus when $T_{\textrm{S}}=0$ the overlap matrices belonging to all pairs of $(k,j)$ are taken into account; 
when $T_{\textrm{S}}=1$ only the diagonal contributions with $k=j$ are included.
 Eq.~(\ref{eqn:screening}) allows us to avoid a less desirable spatial truncation criterion.

Currently, however, the main cost in our OSVMP2 implementation is the integral transformation,
since the full local occupied space is employed throughout transforming complete two-electron integrals.
For example, the first-quarter integral transformation  $(\alpha\beta|\gamma\delta) \rightarrow (i\beta|\gamma\delta)$, 
which scales as $N^5$, limits the size of molecules that we can treat efficiently.
As has been previously demonstrated by Werner and coworkers~\cite{lccsd_1, lmp2}, however,
a linear scaling transformation algorithm can be achieved 
by discarding spatially distant occupied orbital pairs and exploiting integral prescreening techniques.
DF/RI techniques in local approximations~\cite{dfr12int,dfmp2} can further decrease the cost of integral transformations 
by 1--2 orders of magnitude for large molecules.
These techniques will be incorporated into our algorithm in future work.

   \section{Comparing the $\textrm{d}$OSVMP2 and OSVMP2 approximations}

We have introduced two related factorizations of MP2 theory, the direct orbital specific virtual approximation (dOSVMP2)
and the full orbital specific virtual approximation (OSVMP2). In addition, in  dOSVMP2 we can define the amplitudes through
two different residual equations: one obtained via the Hylleraas functional (dOSVMP2), and one obtained by projection (dOSVMP2-P).
We now assess the numerical behaviour of these different schemes for correlation energies and reaction energies.

\subsection{Correlation energies}

The number of OSVs per occupied orbital (the number of $\mu_i$ or $\nu_i$ in e.g. Eq. (\ref{eqn:fac})) 
needed to recover a given accuracy in the correlation energy relative to the canonical MP2 energy
is reported  for  polyglycine oligopeptides, water clusters and polyene chains. 
As seen in Table~\ref{tab:table1}, 
a small number of OSVs  recovers most of the correlation energy (e.g., $\ge 99.5\%$),
the precise number depending on the electronic structure of the molecule.
The full OSVMP2 (which includes the exchange excitations) requires far fewer OSVs to reach
the same accuracy than the direct dOSVMP2, typically less than half. Nonetheless both approximations
are very compact.
The number of OSVs to reach a given accuracy also becomes independent of the total molecular size very rapidly.
For OSVMP2 the correlation energy is  saturated at accuracies of 99.5\%, 99.9\% and 99.99\%, respectively, 
with 20, 29 and 47 orbitals for [gly]$_n$ and with 13, 18 and 29 orbitals for (H$_2$O)$_n$.
Note that this saturation behaviour is expected when the system size becomes much larger than its correlation length.

Compared to [gly]$_n$ and (H$_2$O)$_n$,
the polyene molecule exhibits significant electronic delocalization, 
and this leads to longer correlation lengths and more extended orbitals.
As a result, it is more difficult to converge the correlation energy towards the canonical limit than in other molecules.
For example, with 40 OSVs the OSVMP2 error increases from 0.46\% to 0.66\% (cf. Table~\ref{tab:table2}) 
as the length of polyene chain increases from C$_6$H$_8$ to C$_{14}$H$_{16}$. Note that this decrease in accuracy is 
physical and not
a failure of extensivity of the theory: as the HOMO-LUMO gap of the polyenes
decreases with increasing chain length, the correlation length increases.

Regarding the different residual equations for the direct dOSVMP2 factorization,
both yield very similar results.
Using the Hylleraas residual (dOSVMP2), we need a few more OSVs than the projected residual (dOSVMP2-P) to
recover the same accuracy in the correlation energy. For example, 109 and 105 OSVs respectively yield
99.99\% accuracy in the correlation energy for dOSVMP2 and dOSVMP2-P for the largest peptide [gly]$_{14}$; 
for the (H$_2$O)$_{19}$ cluster we require  78 and 73 OSVs to reach the same accuracy.

\subsection{Reaction energies}

Relative energies are the central quantity in chemistry rather than  absolute energies.
We have investigated their accuracy  by computing relative energies for some isomerization reactions
using the different orbital specific virtual approximations.
These reactions were selected in Ref.~\cite{reactions} for the good agreement between the canonical MP2 isomerization energies with triple-$\zeta$ basis sets (cc-pVTZ) and the experimental isomerization energies.
The results of dOSVMP2-P and OSVMP2 computations as well as their deviations from the canonical MP2 values are given in Table~\ref{tab:isoreac}.
The dOSVMP2-P computation using only 10\% of the OSVs gives a MAD (mean absolute deviation) of 2.33 kcal/mol.
Using 20\% of the OSVs drops the MAD to 0.99 kcal/mol which is chemical accuracy.
The MAD is further reduced to 0.07 kcal/mol (almost two orders of magnitude) when 60\% of the OSVs are used. 

\begin{figure*}[ht!]
\includegraphics[width=1.0\linewidth]{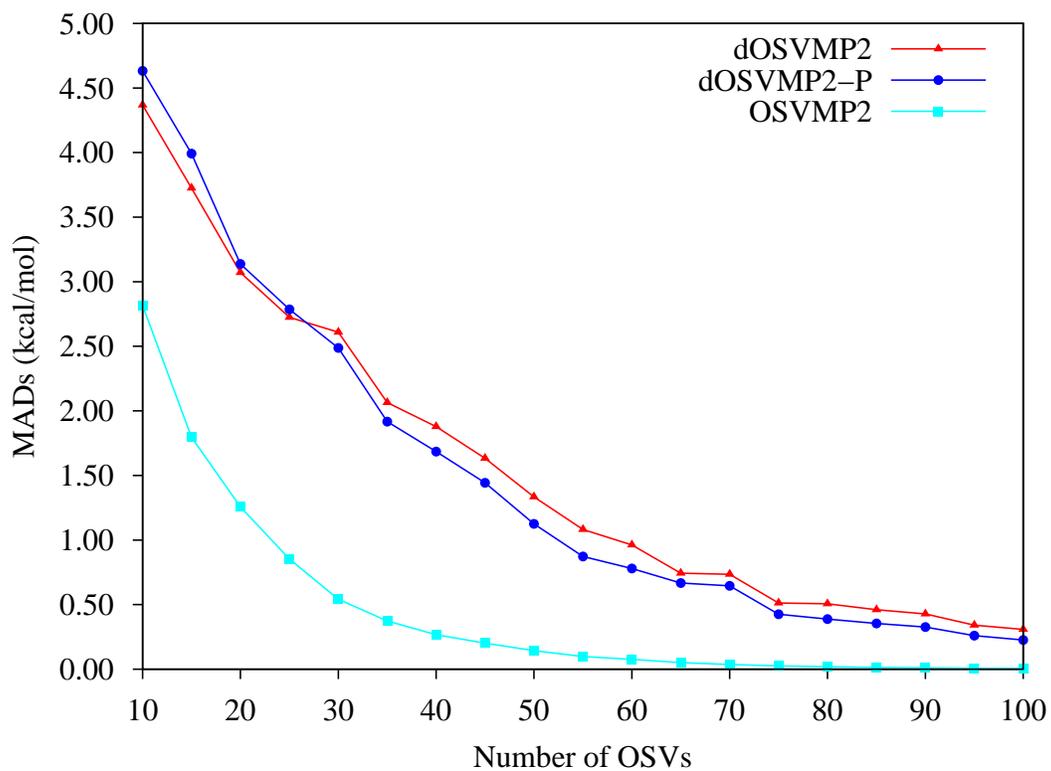} 
\caption{\label{fig:iso} 
         MADs (Mean Absolute Deviations) of the isomerization energies as a function of the number of OSVs using dOSVMP2, dOSVMP2-P and OSVMP2. 
        }
\end{figure*}

The MADs of isomerization energies are plotted against the numbers of OSVs in Fig.~\ref{fig:iso} 
for OSVMP2, dOSVMP2-P and dOSVMP2 schemes. 
Both dOSVMP2-P and dOSVMP2 display similar accuracies. However, 
the complete OSVMP2 gives errors that are substantially smaller and additionally, these errors 
decay more rapidly and more smoothly than those of dOSVMP2 and dOSVMP2-P, as the number of 
OSVs used is increased.
Nonetheless all methods show a rapid decrease in error as the size of the OSV space is increased.

\subsection{Basis set dependence}

\begin{figure*}[ht!]
\includegraphics[width=1.0\linewidth]{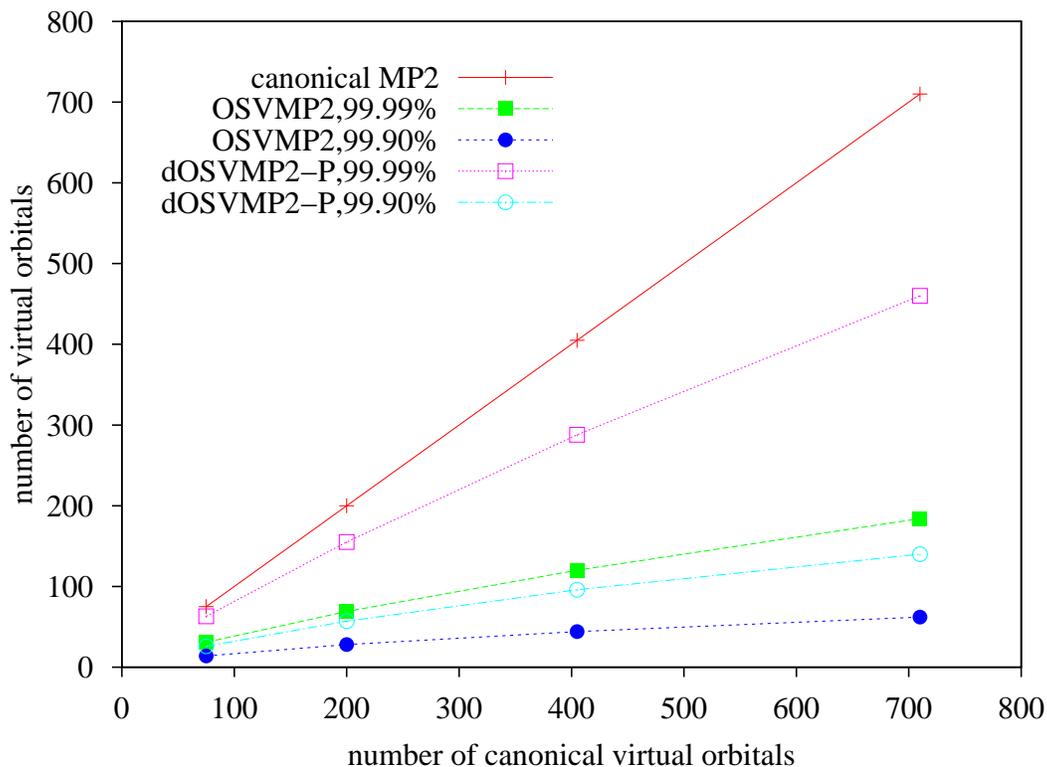} 
\caption{\label{fig:basis} 
      The number of orbital specific and canonical virtual orbitals using different basis sets (cc-pVXZ, X=D, T, Q and 5) for a single glycine molecule.
         Comparison is made between canonical MP2 (+), dOSVMP2-P (unfilled) and OSVMP2 (filled) 
         for accuracies of 99.90\% (circle) and 99.99\% (square) of the correlation energy. Note that OSVMP2 contains
the exchange excitations while the dOSVMP2-P does not. Consequently, the number of virtual orbitals needed for OSVMP2 is significantly
smaller than dOSVMP2, roughly half.
        }
\end{figure*}

We have investigated the orbital specific virtual orbital dependence of different basis sets using the dOSVMP2-P and OSVMP2 approaches. 
For  this  we have chosen to use a single glycine molecule  so that  computations 
with very large basis sets are affordable.
As can be seen from Fig.~\ref{fig:basis}, 
the size of the OSV space (required for a given accuracy in the correlation energy) increases much more slowly than the size of the underlying basis. 
Moving from cc-pVDZ to cc-pV5Z~\cite{dunning},
the size of the required OSV space for dOSVMP2-P and OSVMP2 increases by a factor of 4-5 
while the size of the canonical virtual space increases almost by a factor of 10.
In fact the size of the OSV space needed for a given accuracy appears to increase \textit{sub-linearly} with the size of the underlying basis.

\subsection{Visualizing the orbital specific virtual orbitals}

We have visualized the Boys-localized occupied orbitals~\cite{boys} and a few associated OSVs for
a single glycine molecule in Table~\ref{tab:visorb}.
These local occupied HOMO, HOMO-4 and HOMO-8 orbitals are chosen to be, respectively, 
around the N-lone-pair electrons, C-N and C-C bonds along the skeleton of glycine.
Along each column of Table~\ref{tab:visorb}, each individual OSV exhibits a different shape
for different occupied orbitals.
This can be essentially understood from the definition of the OSVs (cf. Eq.~(\ref{eqn:osv})) since each OSV has to be adjusted to 
a particular occupied orbital. 
For example, the LUMO+4 associated to HOMO, HOMO-4 and HOMO-8
demonstrates, respectively, orbital locality around the N-lone-pair, C-N and C-C bonds,
where the associating occupied orbitals are found.

\section{Comparison with the Pulay-Saeb{\o} local MP2 theory}

The Pulay-Saeb{\o} local correlation approach based on projected atomic orbitals (PAO)
is a standard against which to compare new approaches to local correlation, 
such as the  orbital specific virtual approximations used here. Here we  assess both the accuracy and times of the OSVMP2
and dOSVMP2 approximations relative to the Werner-Sch\"utz formulation of the  Pulay-Saeb{\o} local MP2 
as implemented in \textsc{Molpro} \cite{molpro,lmp2}.

\subsection{Potential energy surfaces}

Pulay-Saeb{\o} local correlation relies on spatial truncation of virtual orbital domains.
Discontinuities on potential energy surfaces (PES) can then arise
since the virtual orbital domain size defined by spatial truncation is not uniform as the geometry is varied.
One prototypical example is the propadienone (CH$_2$CCO) molecule 
that has been recently investigated by Russ et al.~\cite{propadienone} using local CCSD and MP2.
Multiple discontinuities occur in the stretching of the central C=C bond of propadienone, 
even in the vicinity of the equilibrium geometry.
Several attempts have been made to recover smooth PES in Pulay-Saeb{\o} theory.
By tailoring and fixing virtual domains~\cite{discpao1} 
the Pulay-Saeb{\o} local approach can avoid these discontinuities. 
Explicitly correlated R12/F12 methods~\cite{r12f12} reduce the magnitude of discontinuities
through the auxiliary excitation space~\cite{r12domain1,r12domain2}. 
There have also been efforts to use bump functions~\cite{bumpfcn} 
to smooth discontinuous amplitudes.
We now reinvestigate this issue using the orbital specific virtual approximations.
We believe the approach presented here provides a more basic solution.

\begin{figure*}[ht!]
 {\includegraphics[width=0.6\linewidth]{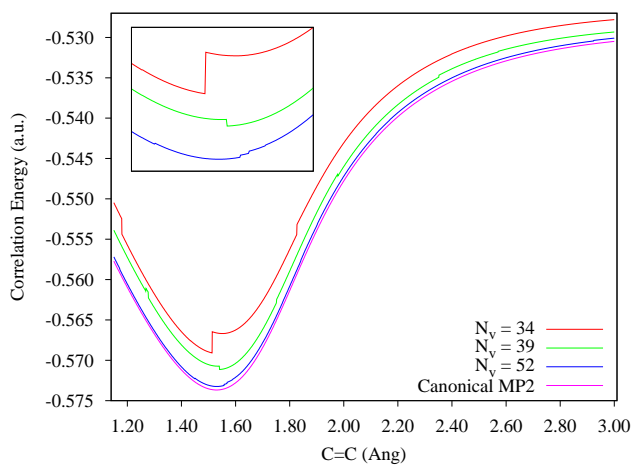}} \\ (a) Pulay-Saeb{\o}'s PAO local MP2  \\[1cm]
 {\includegraphics[width=0.6\linewidth]{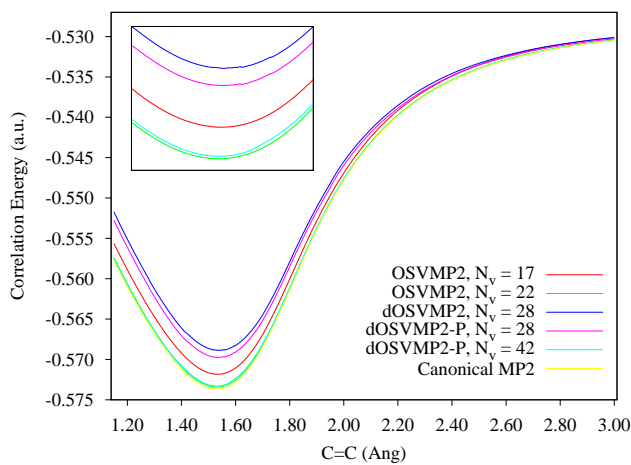}} \\ (b) OSVMP2, dOSVMP2 and dOSVMP2-P   
\caption{\label{fig:pes} 
         Potential energy surfaces resulting from (a) Pulay-Saeb{\o}'s PAO local MP2 (using Molpro~\cite{molpro})
         and (b) OSVMP2, dOSVMP2 and dOSVMP2-P for the central C=C bond of propadienone using a cc-pVDZ basis set~\cite{dunning}.
         The equilibrium geometry was obtained by optimizing all internal coordinates of propadienone at the MP2/cc-pVDZ level.
         The displacement of the central C=C bond was 0.001~\AA~and other internal coordinates were frozen.
         The insets magnify the details at the vicinity of the equilibrium C=C bonds.
        }
\end{figure*}

The correlation energy PES using different numbers of virtual orbitals ($\textrm{N}_\textrm{v}$) are presented in Fig.~\ref{fig:pes} 
for Pulay-Saeb{\o} PAO MP2, dOSVMP2, dOSVMP2-P and OSVMP2. 
All computations were carried out using the cc-pVDZ basis set~\cite{dunning} with 52 canonical virtual orbitals.
For PAO local MP2, $\textrm{N}_\textrm{v}$ denotes the average size of the pair virtual domain.
It can be seen that the PES of the PAO based local MP2 with $\textrm{N}_\textrm{v}=34$ orbitals exhibits three major energy discontinuities
in the regions of both short and long C=C bonds as well as around the equilibrium C=C bond.
When using $\textrm{N}_\textrm{v}=39$ PAOs, five smaller discontinuities in the PAO local MP2 theory appear, 
ranging from 0.2 to 0.4 $mE_h$ at 1.269, 1.540, 1.752, 1.978 and 2.352 \AA.
A tiny zig-zag structure can still be seen in the vicinity of the equilibrium C=C bond even when using $\textrm{N}_\textrm{v}=52$ PAOs 
(see inset of Fig.~\ref{fig:pes}a).

As for the PES using the orbital specific virtual approximations, 
both the dOSVMP2, dOSVMP2-P and OSVMP2 based curves are smooth even when using much smaller spaces of virtual orbitals.
The PES of OSVMP2 displays no discernible discontinuities when using 17 OSVs (cf. Fig.~\ref{fig:pes}b), and
when using 22 OSVs the OSVMP2 based PES is already close to the curve of canonical MP2.
In the case of  dOSVMP2 and dOSVMP2-P with $\textrm{N}_\textrm{v}=28$ OSVs we see very tiny breaks (only 0.04 and 0.02 $mE_h$)
respectively, at 1.559 and 1.597 \AA.
These discontinuities are, nevertheless, one order of magnitude smaller than 
those of the PAO $\textrm{N}_\textrm{v}=34$ result discontinuities.

\subsection{Virtual space size and timings}

\begin{figure*}[ht!]
\includegraphics[width=1.0\linewidth]{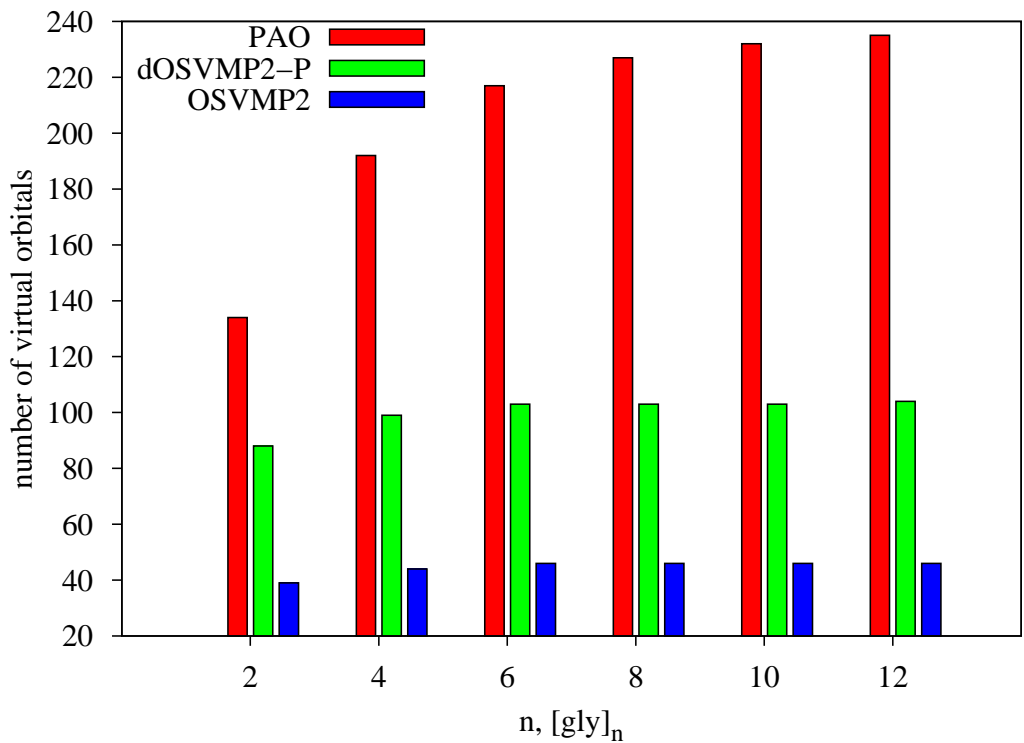} 
\caption{\label{fig:vcmp} 
      Histograms of virtual orbital numbers for Pulay-Saeb{\o}'s PAO local MP2, dOSVMP2-P and OSVMP2 schemes needed to obtain $99.99\%$ accuracy.
      In the case of PAO local MP2, this number is obtained from  the average pair domain size. Note that OSVMP2 contains
the exchange excitations while the dOSVMP2-P does not. Consequently, the number of virtual orbitals needed for OSVMP2 is significantly
smaller than dOSVMP2, roughly half.
        }
\end{figure*}

The  efficiency of both the Pulay-Saeb{\o} PAO and orbital specific virtual approximations
depends on the size of the virtual space needed to obtain good agreement with the canonical result.
Fig.~\ref{fig:vcmp} gives the comparison of virtual space sizes between PAO local MP2, dOSVMP2-P and OSVMP2 needed to
recover 99.99\% of the correlation energy in [gly]$_n$.
It is evident that PAO based local MP2 needs substantially more virtual orbitals (two times) than dOSVMP2-P 
and at least four times more than OSVMP2.
Furthermore, the sizes of the orbital specific virtual spaces for dOSVMP2-P and OSVMP2  saturate much more rapidly than those of PAO based MP2
when the molecular size increases.
The relative advantage of the orbital specific scheme increases as we move to the large basis i.e. from cc-pVDZ to cc-pVTZ~\cite{dunning}.

However, the $O^3$ scaling term in the orbital specific virtual approximations has a higher prefactor than 
that of the similar term in the Pulay-Saeb{\o} PAO theory.
This affects adversely the CPU times to solve the residual equations  when $O$ becomes large, 
even though the orbital specific virtual space  is much smaller than the PAO space used in the Pulay-Saeb{\o} scheme.
For instance, going from [gly]$_4$ to [gly]$_{12}$, for an accuracy of 99.99\%
the CPU times of the projective dOSVMP2-P ($T_\textrm{S}=0$ of Table~\ref{tab:screening1}) start to become less favourable than that of the
Pulay-Saeb{\o} theory.
The CPU time for the full OSVMP2 computations on  [gly]$_{12}$ is  longer than that for the  Pulay-Saeb{\o} implementation
 by a factor of three.

The computational efficiency of the orbital specific virtual approximations can be greatly improved, 
however, by using the screening scheme discussed earlier.
We have  reinvestigated the CPU time to solve the residual equations using screening and the results are presented 
in Table~\ref{tab:screening1} for [gly]$_n$ and Table~\ref{tab:screening2} for polyenes, respectively.
The threshold $T_{\textrm{S}}$ is chosen such that there is only a very minor error in the correlation energy, 
e.g., the screening errors are restricted to less than  0.01\% of the correlation energy or only a few tenths of kcal/mol
in the present study.

For the longest [gly]$_{12}$,  dOSVMP2-P with $T_{\textrm{S}}=0.20$ and OSVMP2 with $T_{\textrm{S}}=0.07$ are respectively 
speeded up by a factor of 2-3 and 5 compared to those with $T_{\textrm{S}}=0.00$.
As a result, both screened computations for [gly]$_n$ (with a screening error of $\Delta{\textrm{E}_{\textrm{scrn}}}=0.0001\%$)  
are almost two times faster than the Pulay-Saeb{\o} PAO implementation.
The screened cc-pVTZ results of dOSVMP2-P and OSVMP2 are shown for [gly]$_4$ and [gly]$_6$ in Table~\ref{tab:screening1}.
Clearly larger basis sets increase the efficiency of screened OSVMP2 computations relative to PAO and dOSVMP2-P.
Polyene chains are more difficult cases as the correlating orbitals are more extended along the chain than in [gly]$_n$.
However, if we use a looser accuracy of $\Delta_{\textrm{E}_{corr}}\le 0.01\%$,
we find that the  screened dOSVMP2-P and OSVMP2 computations are still faster 
than the Pulay-Saeb{\o} implementation by a factor of 2 and 5, respectively.

\section{Conclusions}
   In this work we have described the direct orbital specific and full orbital specifc virtual approximations to local
second order M{\o}ller Plesset perturbation theory. These representations of the amplitudes have been expressed
in a general language of tensor factorization  that also encompasses many
other representations  used in electronic structure theory. As we have showed, the orbital specific virtual approximation can lead
to significant advantages, both in more formal behaviour, such as smoothness of potential energy curves, as well as
in practical times and accuracies, as compared to efficient implementations of the local Pulay-Saeb{\o} correlation ansatz. 
As for the direct versus full orbital specific virtual approximations, when screening is used, the full orbital specific
virtual approximation is superior.

There is much  to be done along the directions of this work. For example our algorithms are not linearly scaling,
because we have not investigated  efficient representations of the occupied space. 
Furthermore, we expect that significant advantages can be had when 
applying orbital specific virtual type approximations to high body excitations. 
We conclude by recognizing that the space of tensor factorizations is very large, with  much remaining to be explored.

\newpage
\begin{acknowledgments}
Y.K. was supported by Ministry of Education, Culture, Sports, Science and Technology-Japan (MEXT) Grant-in-Aid for Young Scientists (B) 21750028.
He also thanks the Core Research for Evolutional Science and Technology Program, 
“High Performance Computing for Multi-Scale and Multi-Physics Phenomena” of the Japan Science 
and Technology Agency for the financial support for his visiting research at University of Bristol.
Early work on this project was performed while F.R.M. was a visiting scholar at Cornell University.
G.K.C. acknowledges support from the National Science Foundation CAREER Award CHE-0645380.
\end{acknowledgments}

\newpage 
\bibliography{TensorFacMP2}


\newpage

\newpage

\newpage

\newpage

\newpage

\newpage


\newpage
\begin{table*}[htb]
\caption{\label{tab:table1}
      Comparisons of the number of OSVs needed to obtain different accuracies (99.5\%, 99.9\% and 99.99\%) 
      in the MP2 correlation energies using dOSVMP2, dOSVMP2-P and OSVMP2 approximations, respectively.
      The cc-pVDZ basis sets~\cite{dunning} were used.
      The canonical MP2 reference correlation energies were obtained using the Molpro program package~\cite{molpro}.
      $N_{\textrm{v}}$ is the total number of canonical virtual orbitals.
        }
 \begin{ruledtabular}
  \begin{tabular}{l r rrr rrr rrr}
  [gly]$_n$ & & \multicolumn{3}{c}{dOSVMP2} & \multicolumn{3}{c}{dOSVMP2-P} & \multicolumn{3}{c}{OSVMP2} \\
     \cline{3-5} \cline{6-8} \cline{9-11}
  $n$ & $N_{\textrm{v}}$  & 99.5\% & 99.9\% & 99.99\% & 99.5\% & 99.9\% & 99.99\% & 99.5\% & 99.9\% & 99.99\% \\
    \hline
    1 & 75 & 35  & 51  & 65   & 32  & 48  & 63  & 16 & 22 & 31  \\  [-1mm]
    2 &131 & 41  & 62  & 91   & 38  & 58  & 88  & 18 & 25 & 39  \\  [-1mm]
    4 &243 & 44  & 68  & 103  & 42  & 65  & 99  & 19 & 28 & 44  \\  [-1mm]
    6 &355 & 46  & 70  & 106  & 43  & 66  & 103 & 19 & 28 & 46  \\  [-1mm]
    8 &467 & 46  & 71  & 108  & 43  & 67  & 103 & 19 & 28 & 46  \\  [-1mm]
   12 &691 & 47  & 71  & 108  & 44  & 68  & 104 & 20 & 29 & 46  \\  [-1mm]
   14 &803 & 47  & 72  & 109  & 44  & 68  & 105 & 20 & 29 & 47  \\  [1mm]
   \hline 
   \hline 
  (H$_2$O)$_n$ & & \multicolumn{3}{c}{dOSVMP2} & \multicolumn{3}{c}{dOSVMP2-P}  & \multicolumn{3}{c}{OSVMP2} \\
     \cline{3-5} \cline{6-8} \cline{9-11}
  $n$ & $N_{\textrm{v}}$ & 99.5\% & 99.9\% & 99.99\% & 99.5\% & 99.9\% & 99.99\% & 99.5\% & 99.9\% & 99.99\% \\
    \hline
   10$_\textrm{prism}$   &190  & 24  & 35  & 62  & 23  & 33  & 59 & 13 & 17 & 25  \\ [-1mm]
   12$_\textrm{Pr444}$   &228  & 24  & 37  & 71  & 24  & 35  & 66 & 13 & 18 & 27  \\ [-1mm]
   14$_\textrm{Pr2444}$  &266  & 24  & 37  & 71  & 24  & 35  & 66 & 13 & 18 & 28  \\ [-1mm]
   16$_\textrm{Pr4444}$  &304  & 25  & 38  & 75  & 24  & 36  & 70 & 13 & 18 & 28  \\ [-1mm]
   18$_\textrm{Pr44244}$ &342  & 25  & 38  & 75  & 24  & 36  & 70 & 13 & 18 & 28  \\ [-1mm]
   19$_\textrm{globular}$&361  & 25  & 39  & 78  & 24  & 36  & 73 & 13 & 18 & 29 
  \end{tabular}
 \end{ruledtabular}
\end{table*}

\newpage
\begin{table*}[htb]
\caption{\label{tab:table2}
      Comparison of the relative errors of dOSVMP2, dOSVMP2-P and OSVMP2 correlation energies for polyenes 
      using different numbers of virtual orbitals (40, 80, 100 and 140).
      cc-pVTZ basis sets were used.
      The canonical MP2 reference correlation energies were obtained using the Molpro program package~\cite{molpro}.
      $N_{\textrm{v}}$ is the total number of canonical virtual orbitals.
        }
 \begin{ruledtabular}
  \begin{tabular}{l r ccc ccc cc}
         & & \multicolumn{3}{c}{dOSVMP2} & \multicolumn{3}{c}{dOSVMP2-P} & \multicolumn{2}{c}{OSVMP2} \\
     \cline{3-5} \cline{6-8} \cline{9-10}
 Polyenes & $N_{\textrm{v}}$ & 80 & 100 & 140 & 80 & 100 & 140 & 40 & 80 \\
    \hline
 C$_6$H$_8$       & 270 & 0.80\% & 0.39\% & 0.10\% & 0.67\% & 0.31\% & 0.07\% & 0.46\% & 0.02\% \\  [-1mm]
 C$_8$H$_{10}$    & 351 & 0.99\% & 0.52\% & 0.17\% & 0.85\% & 0.42\% & 0.13\% & 0.54\% & 0.03\% \\  [-1mm]
 C$_{10}$H$_{12}$ & 432 & 1.12\% & 0.61\% & 0.22\% & 0.97\% & 0.51\% & 0.17\% & 0.59\% & 0.03\% \\  [-1mm]
 C$_{12}$H$_{14}$ & 513 & 1.21\% & 0.68\% & 0.26\% & 1.06\% & 0.57\% & 0.21\% & 0.63\% & 0.03\% \\  [-1mm]
 C$_{14}$H$_{16}$ & 594 & 1.29\% & 0.73\% & 0.29\% & 1.12\% & 0.62\% & 0.23\% & 0.66\% & 0.04\% 
  \end{tabular}
 \end{ruledtabular}
\end{table*}

\newpage
\begin{table*}[htb]
\caption{\label{tab:isoreac}
         Calculated canonical MP2, dOSVMP2-P and OSVMP2 isomerization reaction energies (kcal/mol). 
         The reactions and corresponding molecular geometries were taken from~\cite{reactions}
         The dOSVMP2-P and OSVMP2 results are given as the first and second row for each reaction.
         The errors given in the parentheses are the deviations relative to canonical MP2 reaction energies.
         The Ahlrichs-TZV~\cite{Alhrichs-VTZ, Alhrichs-TZV} basis sets augmented by the cc-pVTZ polarization functions (2d1f, 2p1d)~\cite{dunning} 
         were used together with based on the frozen-core approximation.
         The percentage is the first row indicating the fraction (\%) of the orbital specific virtual space used in the calculation.
        }
 \begin{ruledtabular}
  \begin{tabular}{crrrrr}
    Reactions   &MP2  & \multicolumn{1}{c}{10\%}&  \multicolumn{1}{c}{20\%} 
                &\multicolumn{1}{c}{40\%} &\multicolumn{1}{c}{60\%} \\
    \cline{1-1} \cline{2-2} \cline{3-3} \cline{4-4} \cline{5-5} \cline{6-6} 
    7  & 9.26   &10.75 ( 1.49)& 11.10 ( 1.84)& 10.06 ( 0.80)& 9.46 ( 0.20) \\ [-2mm]
       &        &10.54 ( 1.28)&  9.76 ( 0.50)&  9.28 ( 0.02)& 9.26 ( 0.00) \\ [-1mm]
    8  & 22.20  & 20.63 (-1.57)& 21.30 (-0.90)& 21.82 (-0.38)&22.10 (-0.10) \\ [-2mm]
       &        & 21.29 (-0.91)& 22.00 (-0.20)& 22.19 (-0.01)&22.20 ( 0.00) \\ [-1mm]
    9  & 6.96   &  5.17 (-1.79)&  5.74 (-1.22)&  6.92 (-0.04)& 6.94 (-0.02) \\ [-2mm]
       &        &  6.54 (-0.42)&  6.91 (-0.05)&  6.96 ( 0.00)& 6.96 ( 0.00) \\ [-1mm]
   12  & 47.13  & 37.61 (-9.52)& 47.25 ( 0.12)& 47.01 (-0.12)&47.09 (-0.04) \\ [-2mm]
       &        & 45.76 (-1.37)& 47.03 (-0.10)& 47.12 (-0.01)&47.12 ( 0.00) \\ [-1mm]
   18  & 11.52  & 10.75 (-0.77)& 11.13 (-0.39)& 11.53 ( 0.01)&11.52 ( 0.00) \\ [-2mm]
       &        & 11.28 (-0.24)& 11.48 (-0.04)& 11.52 ( 0.00)&11.52 ( 0.00) \\ [-1mm]
   21  & 1.08   & 1.27  ( 0.19)& 1.03  (-0.05)& 1.11  ( 0.03)& 1.07 (-0.01) \\ [-2mm]
       &        & 1.15  ( 0.07)& 1.09  ( 0.01)& 1.08  ( 0.00)& 1.08 ( 0.00) \\ [-1mm]
   24  & 12.56  & 13.44 ( 0.88)& 12.85 ( 0.29)& 12.36 (-0.20)&12.48 (-0.08) \\ [-2mm]
       &        & 12.68 ( 0.12)& 12.52 (-0.04)& 12.56 ( 0.00)&12.56 ( 0.00) \\ [-1mm]
   28  & 31.12  & 33.90 ( 2.78)& 33.10 ( 1.98)& 31.74 ( 0.62)&31.32 ( 0.20) \\ [-2mm]
       &        & 32.64 ( 1.52)& 31.45 ( 0.33)& 31.14 ( 0.02)&31.12 ( 0.00) \\ [-1mm]
   32  & 7.35   & 3.04  (-4.31)& 5.70  (-1.65)&  7.01 (-0.34)& 7.29 (-0.06) \\ [-2mm]
       &        & 6.20  (-1.15)& 7.15  (-0.20)&  7.35 ( 0.00)& 7.35 ( 0.00) \\ [-1mm]
   34  & 6.98   & 7.00  ( 0.02)& 5.52  (-1.46)&  6.71 (-0.27)& 6.97 (-0.01) \\ [-2mm]
       &        & 6.62  (-0.36)& 6.96  (-0.02)&  6.98 ( 0.00)& 6.98 ( 0.00) \\ 
  \hline
\textbf{M}ean \textbf{A}bsolute \textbf{D}eviation& (kcal/mol)   & \multicolumn{1}{c}{2.33}  &  \multicolumn{1}{c}{0.99}   
                          &  \multicolumn{1}{c}{0.28}     &  \multicolumn{1}{c}{0.07}      \\ [-2mm]
                          &      & \multicolumn{1}{c}{0.74}  &  \multicolumn{1}{c}{0.15}   
                          &  \multicolumn{1}{c}{0.01}     &  \multicolumn{1}{c}{0.00}      \\ [-1mm]
\textbf{M}ean \textbf{D}eviation& (kcal/mol) & \multicolumn{1}{c}{-1.26} &  \multicolumn{1}{c}{-0.14}   
                          &  \multicolumn{1}{c}{0.01}      &  \multicolumn{1}{c}{0.01}       \\ [-2mm]
                          &      & \multicolumn{1}{c}{-0.15} &  \multicolumn{1}{c}{0.02}   
                          &  \multicolumn{1}{c}{0.00}      &  \multicolumn{1}{c}{0.00}       \\ [-1mm]
\textbf{M}aximum \textbf{D}eviation & (kcal/mol)  &  \multicolumn{1}{c}{9.52}&  \multicolumn{1}{c}{1.98}     
                          &  \multicolumn{1}{c}{0.80}    &   \multicolumn{1}{c}{0.20}        \\ [-2mm]
                          &      &  \multicolumn{1}{c}{1.52}&  \multicolumn{1}{c}{0.50}     
                          &  \multicolumn{1}{c}{0.02}    &   \multicolumn{1}{c}{0.00}      
  \end{tabular}
 \end{ruledtabular}

\end{table*}

\newpage
\begin{table*}[htb]
\caption{\label{tab:visorb}
          Contour plots of a few Boys-localized occupied orbitals (HOMO, HOMO-4 and HOMO-8) 
          and associated OSVs (LUMO, LUMO+4 and LUMO+5) for an isolated single glycine molecule. 
          The OSVs are sorted in  descending order according to the singular values (cf. Eq.~(\ref{eqn:tau})).
          HOMO and LUMO correspond to the orbital, respectively, that has the highest occupied orbital energy 
          and the largest singular value.
          HOMO-4 and HOMO-8 give the fourth and eighth localized occupied orbitals with orbital energies
          below the HOMO.
          LUMO+4 and LUMO+5 are the fourth and fifth OSVs with the singular values above the LUMO.
        }
 \begin{ruledtabular}
  \begin{tabular}{c|cccc}
  \includegraphics[width=0.04\linewidth]{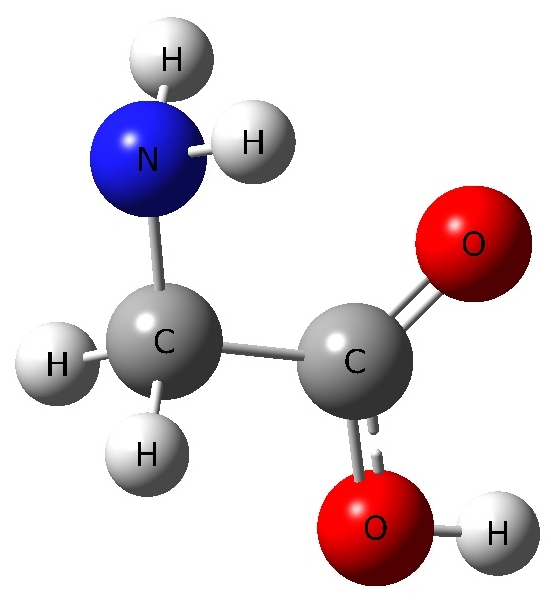} & Local Occ. & LUMO & LUMO+4 & LUMO+5 \\
  \hline\\
  \raisebox{2ex}{HOMO} & \includegraphics[width=0.08\linewidth]{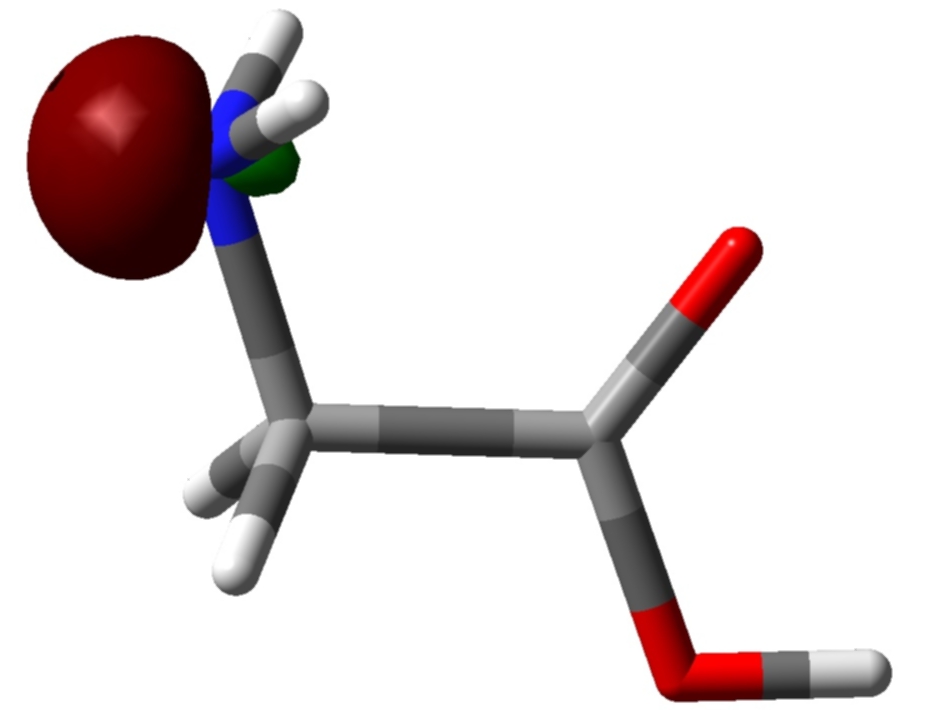} 
          & \includegraphics[width=0.08\linewidth]{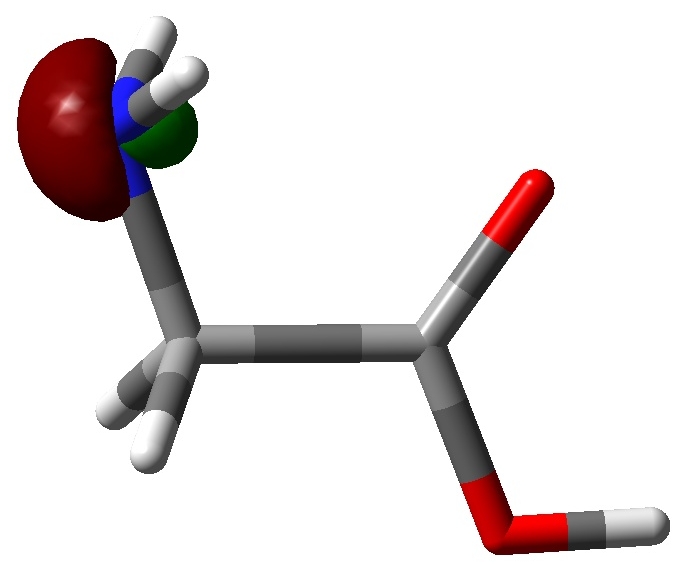} 
                  & \includegraphics[width=0.08\linewidth]{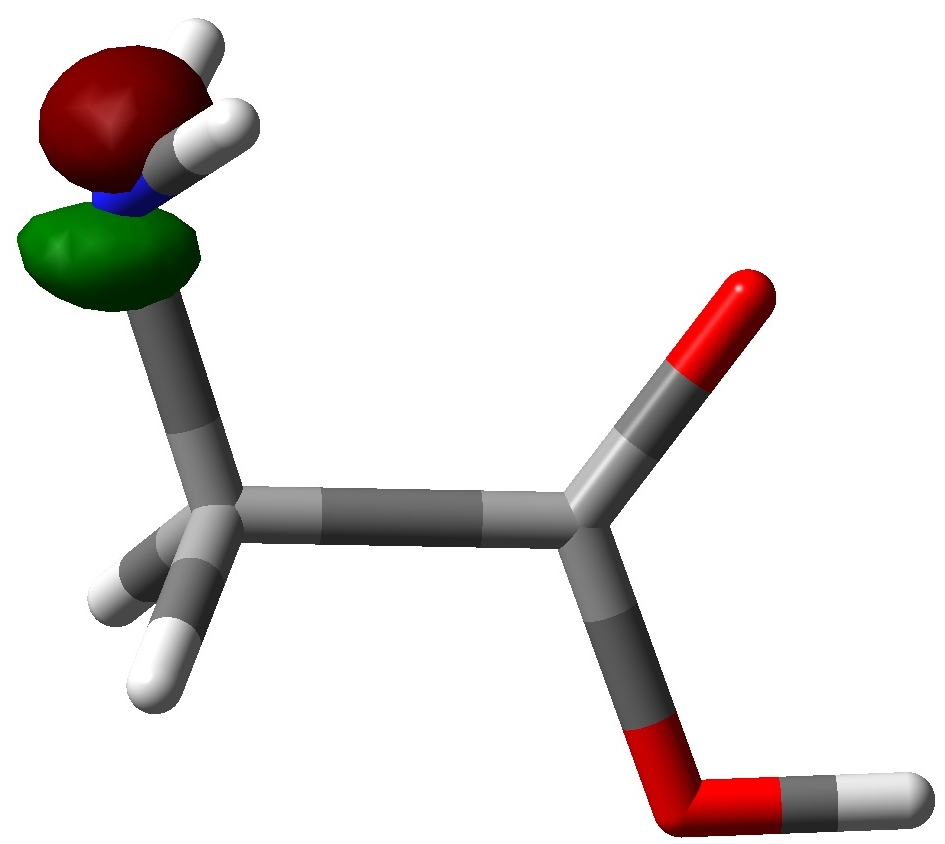} 
                       &  \includegraphics[width=0.08\linewidth]{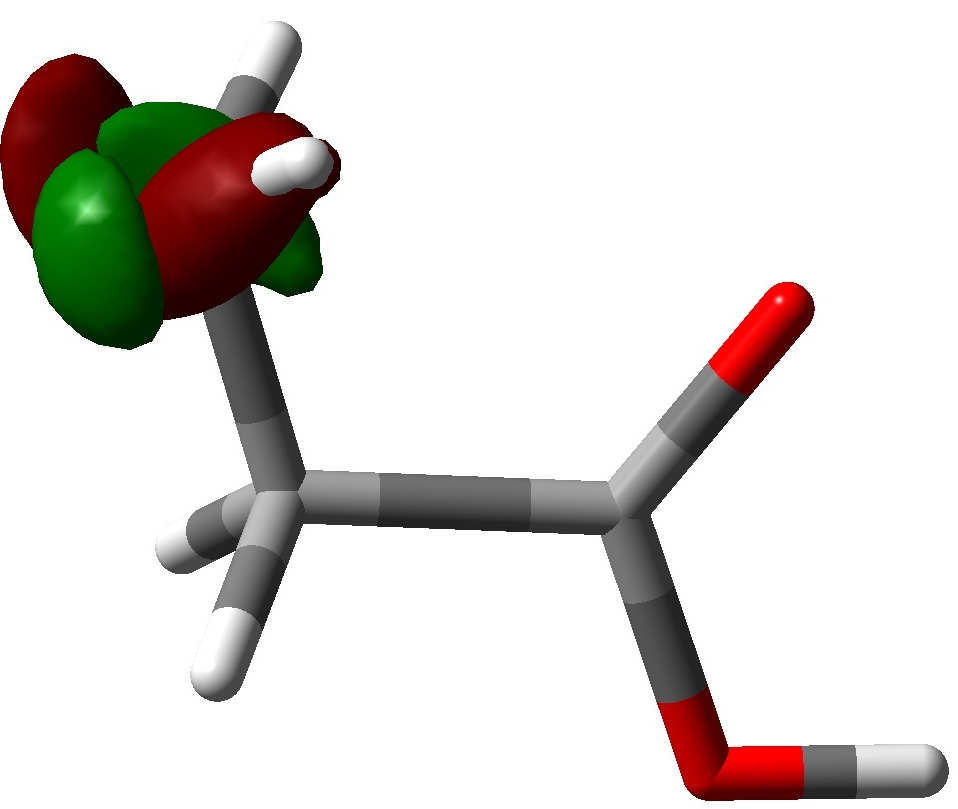} \\   
  \raisebox{2ex}{HOMO-4} & \includegraphics[width=0.08\linewidth]{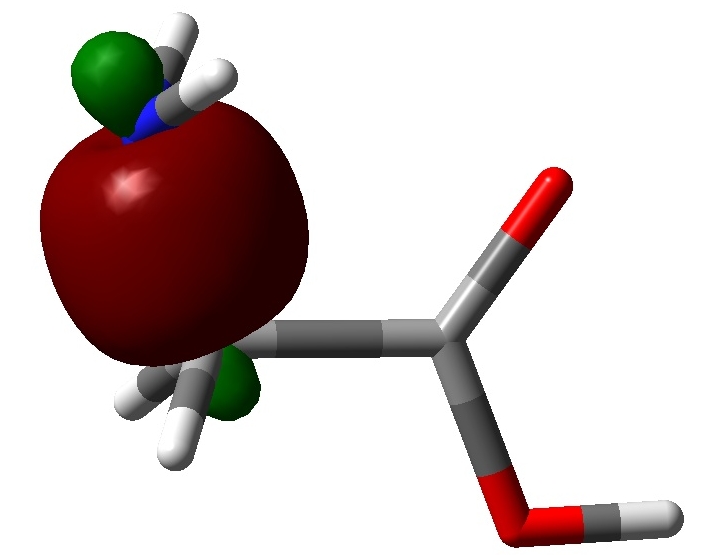} 
             & \includegraphics[width=0.08\linewidth]{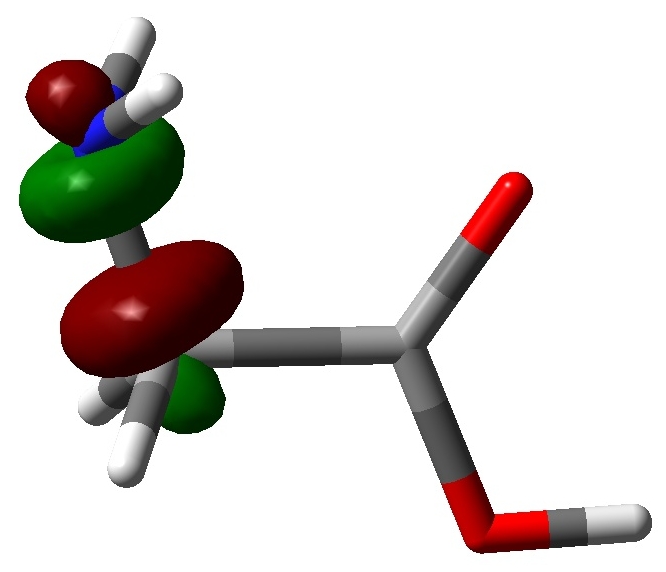} 
                  & \includegraphics[width=0.08\linewidth]{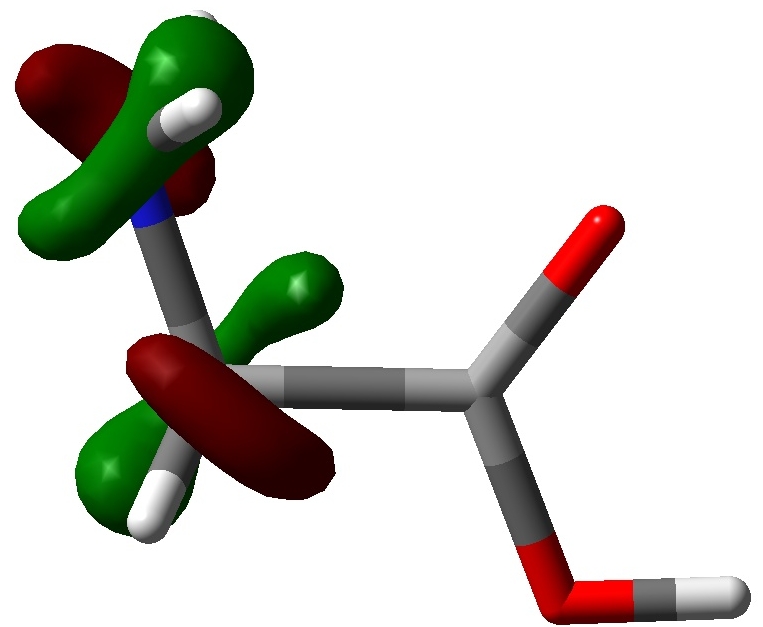} 
                       &  \includegraphics[width=0.08\linewidth]{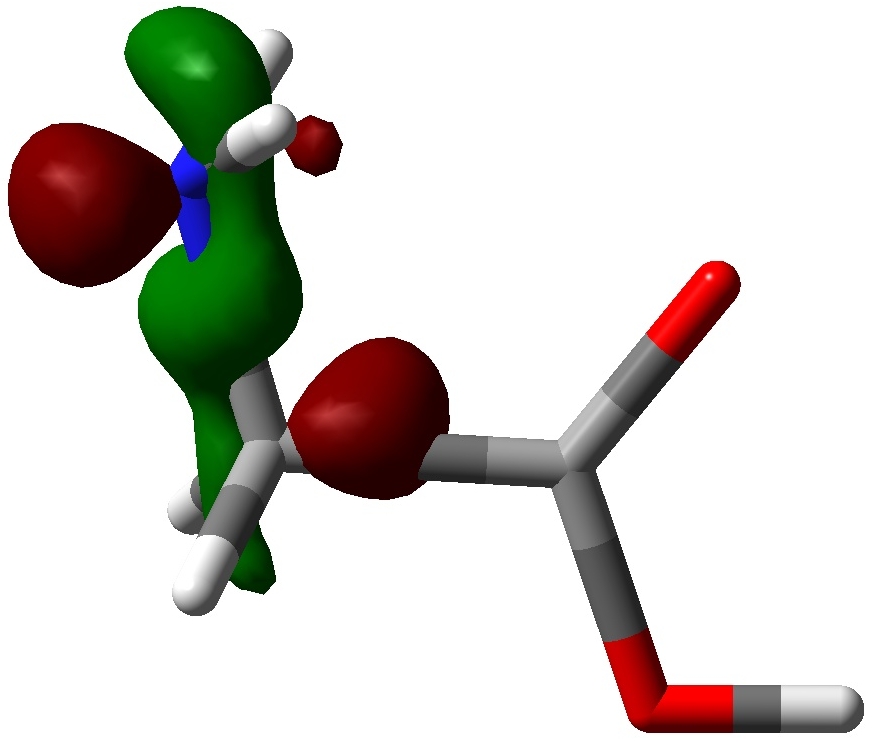} \\   
  \raisebox{2ex}{HOMO-8} & \includegraphics[width=0.08\linewidth]{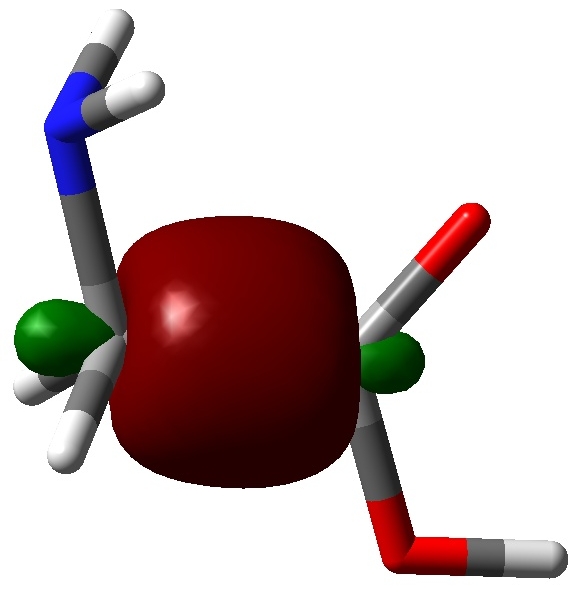} 
             & \includegraphics[width=0.08\linewidth]{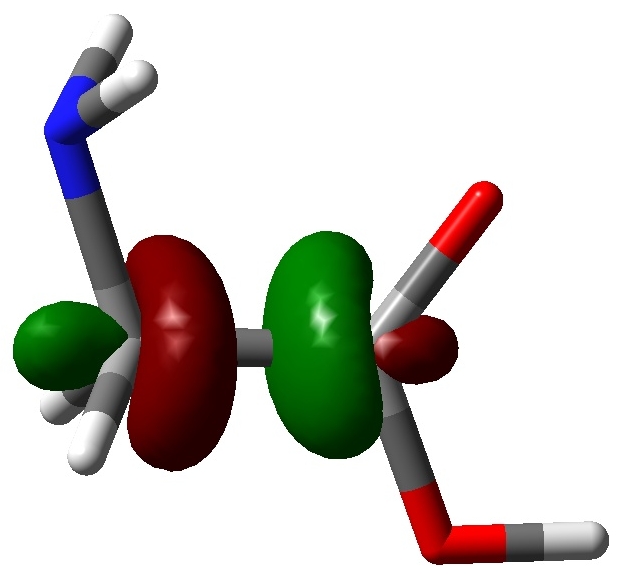} 
                  & \includegraphics[width=0.08\linewidth]{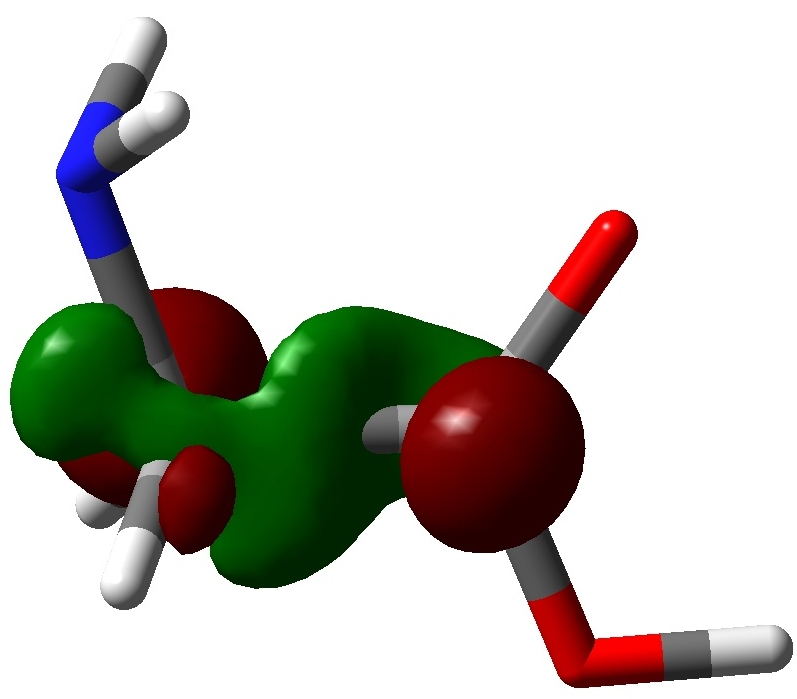} 
                       &  \includegraphics[width=0.08\linewidth]{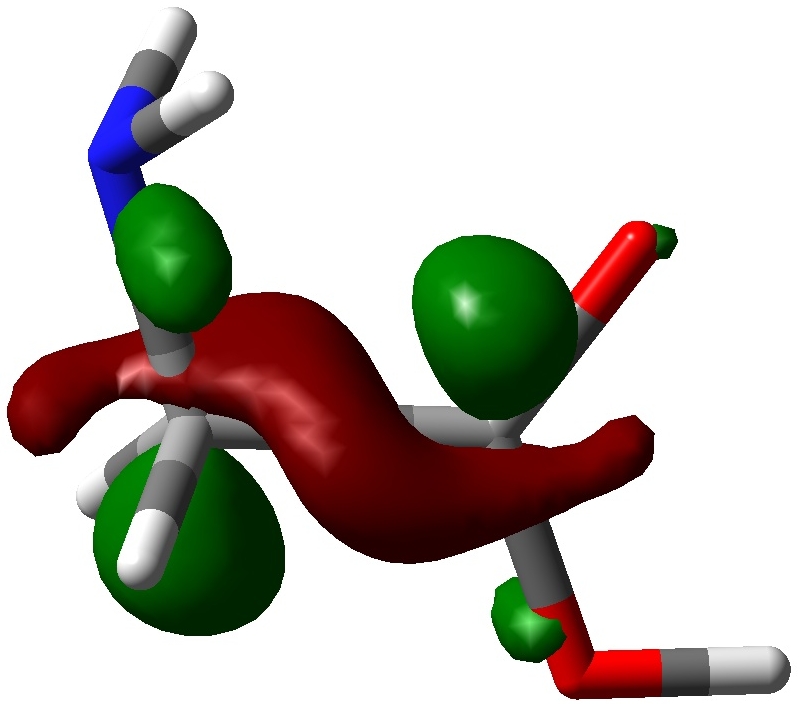} \\   
  \end{tabular}
 \end{ruledtabular}
\end{table*}
\newpage

\begin{table*}[htb]
\caption{\label{tab:screening1}  
         CPU times ($t_{\textrm{solv}}$, seconds) to solve the dOSVMP2-P and OSVMP2 residual equations,
         using different screening thresholds $T_{\textrm{S}}$ 
        for [gly]$_n$ chains with $n=4, 6, 8, 10, 12$.
        103 and 46 OSVs have been used respectively for dOSVMP2-P and OSVMP2 in order to obtain accuracies of 99.99\% 
        for all $T_{\textrm{S}}=0.00$ computations.
        $\Delta{\textrm{E}_{\textrm{scrn}}}$ is the additional relative error in the correlation energy introduced by screening $T_{\textrm{S}}>0.00$
        as compared to canonical MP2. 
        The cc-pVDZ basis set~\cite{dunning} (denoted as D) was used for all computations 
        and results with the cc-pVTZ basis set (denoted as T) are also reported for [gly]$_4$ and [gly]$_6$ molecules.
        With the cc-pVTZ basis set
        $N_v=235$ ([gly]$_4$) and $N_v=243$ ([gly]$_6$) were used for dOSVMP2-P while $N_v=92$ ([gly]$_4$) and $N_v=94$ ([gly]$_6$) were used for OSVMP2.
        } 
\begin{ruledtabular}
\begin{tabular}{r|c|ccccc|ccccc}
  &     & \multicolumn{5}{c|}{$N_v=103$ for dOSVMP2-P} & \multicolumn{5}{c}{$N_v=46$ for OSVMP2} \\
  \cline{3-7} \cline{8-12}
  & PAO & $T_{\textrm{S}}=0.00$ & \multicolumn{2}{c}{$T_{\textrm{S}}=0.20$} & \multicolumn{2}{c|}{$T_{\textrm{S}}=0.40$} 
        & $T_{\textrm{S}}=0.00$ & \multicolumn{2}{c}{$T_{\textrm{S}}=0.07$} & \multicolumn{2}{c}{$T_{\textrm{S}}=0.20$} \\
   \cline{2-2} \cline{3-3} \cline{4-5} \cline{6-7} \cline{8-8} \cline{9-10}  \cline{11-12}
 [gly]$_n$ & $t_{\textrm{solv}}$ 
   & $t_{\textrm{solv}}$ & $t_{\textrm{solv}}$ & $\Delta{\textrm{E}_{\textrm{scrn}}}$ & $t_{\textrm{solv}}$ & $\Delta{\textrm{E}_\textrm{{scrn}}}$ 
   & $t_{\textrm{solv}}$ & $t_{\textrm{solv}}$ & $\Delta{\textrm{E}_{\textrm{scrn}}}$ & $t_{\textrm{solv}}$ & $\Delta{\textrm{E}_\textrm{{scrn}}}$ \\
 \hline
 4 (D) &400.8 & 470.1 &  378.1& 0.0001\%  &  340.3 & 0.0002\% & 624.7   & 307.4 & 0.0001\% & 242.5 & 0.008\% \\ 
 4 (T) &4819.4& 5374.8& 4041.2& 0.0001\%  & 3614.2 & 0.0001\% & 4270.1  &1691.6 & 0.0005\% &1414.5 & 0.01\% \\
 6 (D) &1280.6& 1303.5&  832.5& 0.0001\%  &  749.2 & 0.0004\% & 2104.1  & 741.7 & 0.0001\% & 568.9 & 0.009\% \\
 6 (T) &12767 & 16429 & 10032 & 0.0002\%  & 8872.6 & 0.0004\% & 13911   &4015.3 & 0.0005\% &3328.2 &  0.01\% \\
 8 (D) &2436.7& 2748.8& 1465.7& 0.0001\%  & 1309.7 & 0.0004\% & 4739.4  &1303.8 & 0.0001\% & 984.8 & 0.010\% \\
10 (D) &4038.1& 5050.7& 2283.7& 0.0001\%  & 2027.1 & 0.0005\% & 8981.6  &2023.5 & 0.0001\% &1515.1 & 0.011\% \\
12 (D) &5147.0& 8248.9& 3270.5& 0.0001\%  & 2912.2 & 0.0005\% & 15219.3 &2901.4 & 0.0001\% &2153.8 & 0.011\% \\
\end{tabular}
\end{ruledtabular}
\end{table*}

\newpage
\begin{table*}[htb]
\caption{\label{tab:screening2}  
        CPU times ($t_{\textrm{solv}}$, seconds) to solve the dOSVMP2-P and OSVMP2 residual equations,
        using different screening thresholds $T_{\textrm{S}}$,  for polyenes. 
        The  PAO computation recovers about 99.26\% of the correlation energy.
        100 and 40 OSVs have been used for dOSVMP2-P and OSVMP2, respectively. 
        $\Delta{\textrm{E}_{\textrm{scrn}}}$ is the additional relative error of correlation energy introduced by screening $T_{\textrm{S}}>0.00$ 
        compared to canonical MP2. 
        For $T_{\textrm{S}}=0.00$ the percentage ($\Delta{\textrm{E}_{\textrm{corr}}}$) of the canonical MP2 energy recovered is also reported.
        The cc-pVTZ basis set was used for all computations.
        } \vspace{2mm}
\begin{ruledtabular}
\begin{tabular}{lccccccccccc}
  &     & \multicolumn{6}{c}{$N_v=100$ for dOSVMP2-P} & \multicolumn{4}{c}{$N_v=40$ for OSVMP2} \\
  \cline{3-8} \cline{9-12}
  & PAO &\multicolumn{2}{c}{$T_{\textrm{S}}=0.00$} & \multicolumn{2}{c}{$T_{\textrm{S}}=0.10$} & \multicolumn{2}{c}{$T_{\textrm{S}}=0.20$}
        &\multicolumn{2}{c}{$T_{\textrm{S}}=0.00$} & \multicolumn{2}{c}{$T_{\textrm{S}}=0.03$}  \\
   \cline{2-2} \cline{3-4} \cline{5-6} \cline{7-8} \cline{9-10} \cline{11-12}
 Polyenes & $t_{\textrm{solv}}$ 
          & $t_{\textrm{solv}}$ & $\Delta{\textrm{E}_{\textrm{corr}}}$ 
          & $t_{\textrm{solv}}$ & $\Delta{\textrm{E}_{\textrm{scrn}}}$ 
          & $t_{\textrm{solv}}$ & $\Delta{\textrm{E}_{\textrm{scrn}}}$ 
          & $t_{\textrm{solv}}$ & $\Delta{\textrm{E}_{\textrm{corr}}}$ 
          & $t_{\textrm{solv}}$ & $\Delta{\textrm{E}_{\textrm{scrn}}}$  \\
 \hline
 C$_6$H$_8$        &20.4 & 26.2 & 99.61\% &24.3  & 0.001\% &  23.1 & 0.004\% & 15.2 & 99.54\%&11.2  & 0.002\%  \\
 C$_8$H$_{10}$     &53.5 & 50.4 & 99.48\% &43.3  & 0.002\% &  40.2 & 0.007\% & 33.0 & 99.46\%&19.6  & 0.004\%  \\
 C$_{10}$H$_{12}$  &110.3& 86.2 & 99.39\% &68.8  & 0.003\% &  63.4 & 0.009\% & 62.8 & 99.41\%&31.8  & 0.005\%  \\
 C$_{12}$H$_{14}$  &211.5& 135.0& 99.32\% &98.6  & 0.004\% &  89.6 & 0.009\% & 103.4& 99.37\%&44.3  & 0.007\%  \\
 C$_{14}$H$_{16}$  &313.8& 224.4& 99.27\% &137.1 & 0.004\% & 123.2 & 0.009\% & 163.7& 99.34\%&62.1  & 0.008\%  \\
\end{tabular}
\end{ruledtabular}
\end{table*}

\end{document}